\documentclass{aastex631}
\usepackage{float}

\begin{document}

\title{Investigate Substructures of the Inner Halo Using the StarGO Algorithm with APOGEE DR17}

\correspondingauthor{Cuihua Du}
\email{ducuihua@ucas.ac.cn}

\author{Zhongcheng Li}
\affiliation{School of Astronomy and Space Sciences, University of Chinese Academy of Sciences, Beijing 100049, P.R. China}

\author{Cuihua Du}
\affiliation{School of Astronomy and Space Sciences, University of Chinese Academy of Sciences, Beijing 100049, P.R. China}

\author{Haoyang Liu}
\affiliation{School of Astronomy and Space Sciences, University of Chinese Academy of Sciences, Beijing 100049, P.R. China}

\author{Dashuang Ye}
\affiliation{School of Astronomy and Space Sciences, University of Chinese Academy of Sciences, Beijing 100049, P.R. China}



\begin{abstract}
We applied Stars’ Galactic Origin (StarGO) algorithm in 7-D space (i.e. [Fe/H], [Mg/Fe], [Al/Fe],\, $L_{z},\,J_{r},\,J_{z},\,E$) to analyze stars in the inner halo with APOGEE DR17. We identified some known substructures, including Gaia–Sausage–Enceladus (GSE), Sagittarius Stream (Sgr), LMS-1 (Wukong), Thamnos, Metal-Weak Thick Disk (MWTD) and Aleph. Additionally, we identified an undefined metal-poor group (UDG, with [Fe/H] \(< -0.8 \, \)dex) probably linked to known substructure like Aleph, as well as a high \(\alpha\)-abundance substructure (HAS) associated with both the Nyx and Nyx-2 streams. Chemical abundance of the HAS supports the argument that Nyx and Nyx-2 share a common origin. We discovered three substructures, which we refer to as new-substructure candidate-1,2,3 (NSTC-1,2,3). Despite exhibiting disk-like dynamics, these NSTCs demonstrate notably low [Mg/Fe] ($<$ 0.2 dex) and [Al/Fe] ($< -$0.15 dex), similar to the properties of dwarf galaxies.‌ Their high orbital energy and low $[\alpha / \text{Fe}]$ indicate the association with recent accretion events.

\end{abstract}



\section{Introduction} \label{sec:intro}
The study on evolution of the Milky Way (MW) throws light on the galaxy formation and assembly. According to the $\Lambda$CDM model, MW is formed mainly through hierarchical assembly, where nearby satellite galaxies are disrupted and accreted \citep{White1991ApJ}. Since different galaxies have different star formation (SF) histories, accreted galaxies tend to exhibit unique chemical characteristics. Therefore, the evolution history and components of the MW are significantly influenced. As a result, the merger events of different dwarf galaxies leave debris known as ``substructure'' in the MW \citep[e.g.,][]{Ibata1994ads,Helmi1999Debris,Belokurov2018Co,Barb2019as,Koppelman2019Multiple,Necib2020evidence,Yuan2020als,Naidu2020evidence,Newberg2009discovery,Malhan2022the}. Accreted satellites can be effectively recovered in motion space defined by energy $E$, total angular momentum $L$ and its z-component $L_{z}$ due to the conservation of energy and angular momentum \citep{HelmiZeeuw2000MNRAS}. Specifically, $L_{z}$ is conserved under an axisymmetric gravitational potential, while $L$ is conserved under a spherically symmetric potential.
 For example, Thamnos was initially identified in the $E -L_z$ space \citep{Koppelman2019Multiple}, while substructures like Arjuna, Aleph, I'itoi and Sequoia were discovered in action space \citep{Myeong2019evidence,Naidu2020evidence}. Due to the different timescale of SNe I and II events \citep{Timmes1995ApJS}, the chemical properties of these substructures could be explored in the [$\alpha$/Fe]–[Fe/H] plane, as well as through neutron capture processes (r-process and s-process), revealing the star formation history of their progenitors.
\par 
The Sagittarius dwarf galaxy (Sgr) is currently being accreted by the MW, suggesting that it has not yet had sufficient time to be phase-mixed, and its stellar streams remain traceable \citep{Ibata1994ads}. In contrast, the debris from disrupted dwarf galaxies in past merger events is no longer detectable as stream-like structures at present, while the clumps (known as substructures) in kinematic and chemical spaces remain traceable due to the large dynamical timescale. Gaia–Sausage–Enceladus (GSE) is a substructure identified by \citet{Belokurov2018Co} and \citet{Helmi2018the} and it is named for its blob-shaped/sausage-shaped distribution in the $V_{r}$-$V_{\phi}$ space. GSE is thought to be accreted $\sim\,$8 -10 Gyr ago \citep{DiMatteo2019A&A,Gallart2019NatAs} and comprised of the major part ($\sim 41\% - 71\%$) of the inner halo \citep{Wu2022contribution}. Recently, \citet{Ye2024compositions} found that the Hercules–Aquila Cloud (HAC) and Virgo Overdensity (VOD) are dominated by GSE stars with weights of approximately 0.67 and 0.57, respectively. The Helmi Streams were identified in (\( L_z \), \( L_{\perp} \)) space, exhibiting notably high \( V_z \) and \( L_{\perp} \) values, where \( L_{\perp} = \sqrt{L_{x}^{2} + L_{y}^{2}} \) represents the in-plane angular momentum component and \( V_z \) denotes the vertical velocity \citep{Helmi1999Debris}. The Helmi Streams can be divided into two distinct components: one characterized by positive $V_z$ and the other by negative $V_z$ \citep{Koppelman2019characterization}. They are assumed to be accreted from a dwarf galaxy 5 to 8 Gyr ago \citep{Koppelman2019Multiple} with a total mass of $\sim 10^{8}\, \mathrm{M}_{\odot}$ \citep{Naidu2020evidence}. LMS-1 (Wukong) is a less prograde and more metal-poor substructure compared to the Helmi Streams, and it is likely to have originated from a dwarf satellite with a total mass of approximately \( 2 \times 10^{9} M_{\odot} \) \citep{Yuan2020als,Naidu2020evidence}. Thamnos is a retrograde substructure that can be divided into two components, Thamnos 1 and Thamnos 2, based on metallicity and azimuthal velocity, and is likely to have originated from a small dwarf galaxy with a stellar mass of less than \( 5 \times 10^6 M_{\odot} \) \citep{Koppelman2019Multiple}.
 Aleph is discovered by \citet{Naidu2020evidence} and clearly differentiated from typical disk populations for its notable vertical action ($J_{z}$). The MWTD represents the metal-poor tail of the high-\(\alpha\) disk and is likely the result of a merger event or the early evolution of the MW \citep{Carollo2019evidence}. Nyx is a disk substructure firstly considered ex-situ due to its prograde, highly eccentric motion and significant radial velocity component \citep{Necib2020evidence}. However, it is then proved to be most likely a high-velocity component of the Galactic thick disk \citep{Zucker2021ApJ}. Detailed elemental abundances also give an alternative view that Nyx could have an in-situ origin because of similar chemical pattern to that of the thick disk \citep{Wang2023ApJ}, which may result from the spin-up process of the early disk \citep{Belokurov2022}. Other substructures including Sequoia \citep{Barb2019as,Matsuno2019origin,Myeong2019evidence}, I’itoi and Arjuna \citep{Naidu2020evidence}, Icarus \citep{ReFiorentin2021Icarus}, Cetus \citep{Newberg2009discovery,Yuan2019revealing} and Pontus \citep{Malhan2022the} are discovered in recent years and contribute to the study of the merger history of MW.
\par
The advent of large spectroscopic surveys has provided abundant information on kinematics and chemical abundance. Surveys such as Apache Point Observatory Galaxy Evolution Experiment \citep[APOGEE,][]{Abdurro2022ApJthe}, Large Sky Area Multi-Object Fiber Spectroscopy Telescope \citep[LAMOST,][]{Zhao2012RAA,Cui2012RAA}, the Galactic Archaeology with Hermes survey \citep[GALAH,][]{DeSilva2015GALAH}, the Sloan Extension for Galactic Understanding and Exploration survey \citep[SEGUE,][]{Yanny2009segue}, the Radial Velocity Experiment survey \citep[RAVE,][]{Steinmetz2020the}, and Global Astrometric Interferometer for Astrophysics \citep[Gaia,][]{Gaia2016A&A,gaia2023A&A} have collectively contributed to this progress.
\par In this study, we applied APOGEE DR17 to explore substructures through kinematic and elemental abundance analysis, aiming to uncover intricate relations among inner halo substructures. Employing the StarGO algorithm \citep{Yuan2018StarGO}, we conducted an in-depth investigation of the kinematic and chemical spaces of Galactic inner halo. In Section 2, we describe the data and selection criteria. In Section 3, we present the clustering methodology. The identified substructures are analyzed in Section \ref{4}. Finally, a summary of our work is provided in Section \ref{5}.

\section{Data} \label{2}
In this work, we use the APOGEE DR17 \citep{Abdurro2022ApJthe} to extract kinematic and chemical characteristics of the inner halo. We match \text{the\,astroNN\,catalog\,of\,abundances,\,distances,\,and\,ages\,for\,APOGEE\,DR17\,}stars provided by \citet{Mackereth2019} and \citet{Leung2019}, for detailing distances, ages, and orbital properties. These parameters are derived using the \href{https://github.com/henrysky/astroNN}{astroNN} \citep{Leung20181} deep-learning framework and the fast method of \citet{Mackereth2018PASP}, adopting {{\tt\string MWPotential2014}} gravitational potential \citep{Bovy2015}, the distance of the Sun to the GC is 8.125 kpc \citep{GRAVITYCollaboration2018A&A}, a solar offset of 20.8 pc \citep{Bennett2018}, and solar motion $[U_{\odot} ,V_{\odot} ,W_{\odot} ] = [11.1, 12.24, 7.25]\, \mathrm{km\,s^{-1}}$ \citep{sch2010MNRAS}. Additionally, a chemical abundance catalogue ( $\href{https://www.sdss4.org/dr17/irspec/caveats#wronglsf}{\mathrm{\texttt{allStar}\,catalogue}}$) is utilized. To ensure the data quality, we initially define a parent sample based on the following selection criteria:
\par $\bullet$\, 4000$<T_{\mathrm{eff}}\,<$6000 (K)
\par $\bullet\, $ \texttt{AL\_FE\_FLAG} = 0
\par $\bullet\, $ \texttt{SNREV }$>$ 70
\par $\bullet\, $\texttt{STARFLAG} $\neq$ \texttt{VERY\_BRIGHT\_NEIGHBOR} and \texttt{PERSIST\_HIGH}
\par $\bullet\, $\texttt{ASPCAPFLAG} $\neq$ \texttt{STAR\_BAD}, \texttt{CHI2\_BAD}, \texttt{M\_H\_BAD} and \texttt{CHI2\_WARN}
\par $\bullet \, $\texttt{EXTRATARG} $< $16
\par $\bullet\,\mathrm{log}\,g <\,3.6$
\par $\bullet\,\frac{\mathrm{distance\_error}}{\mathrm{distance}}\leqslant0.2 $

\par The application of effective temperature threshold minimizes systematic effects (e.g., red giant stars with $T_{\mathrm{eff}}\,<$ 4000 K may preferentially have discrete abundance values, leading to unreliable Al abundance measurements). The SNREV criterion ensures the high quality of chemical abundance measurements, while a log $g$ threshold minimizes contamination from dwarf stars. Additionally, the EXTRATARG criterion is implemented to remove duplicated data. The last criteria is applied to ensure relatively accurate distance estimates. Possible Globular and Open Cluster member stars are excluded by removing all the stars contained in catalogues provided by \citet{Schiavon2024} and \citet{Donor2020}. The Large Magellanic Cloud (LMC) and Small Magellanic Cloud (SMC) are excluded according to their galactic longitude and latitude by applying cuts on angular distance with 15$^{\circ }$ and 10$^{\circ }$, from their respective centers, respectively. By applying the above selection criteria, we obtain a parent sample of 222,388 stars. Unless otherwise specified, gray dots in following Figures in this work represent the parent sample stars.

\begin{figure}
    \centering
    \includegraphics[width=0.5\linewidth]{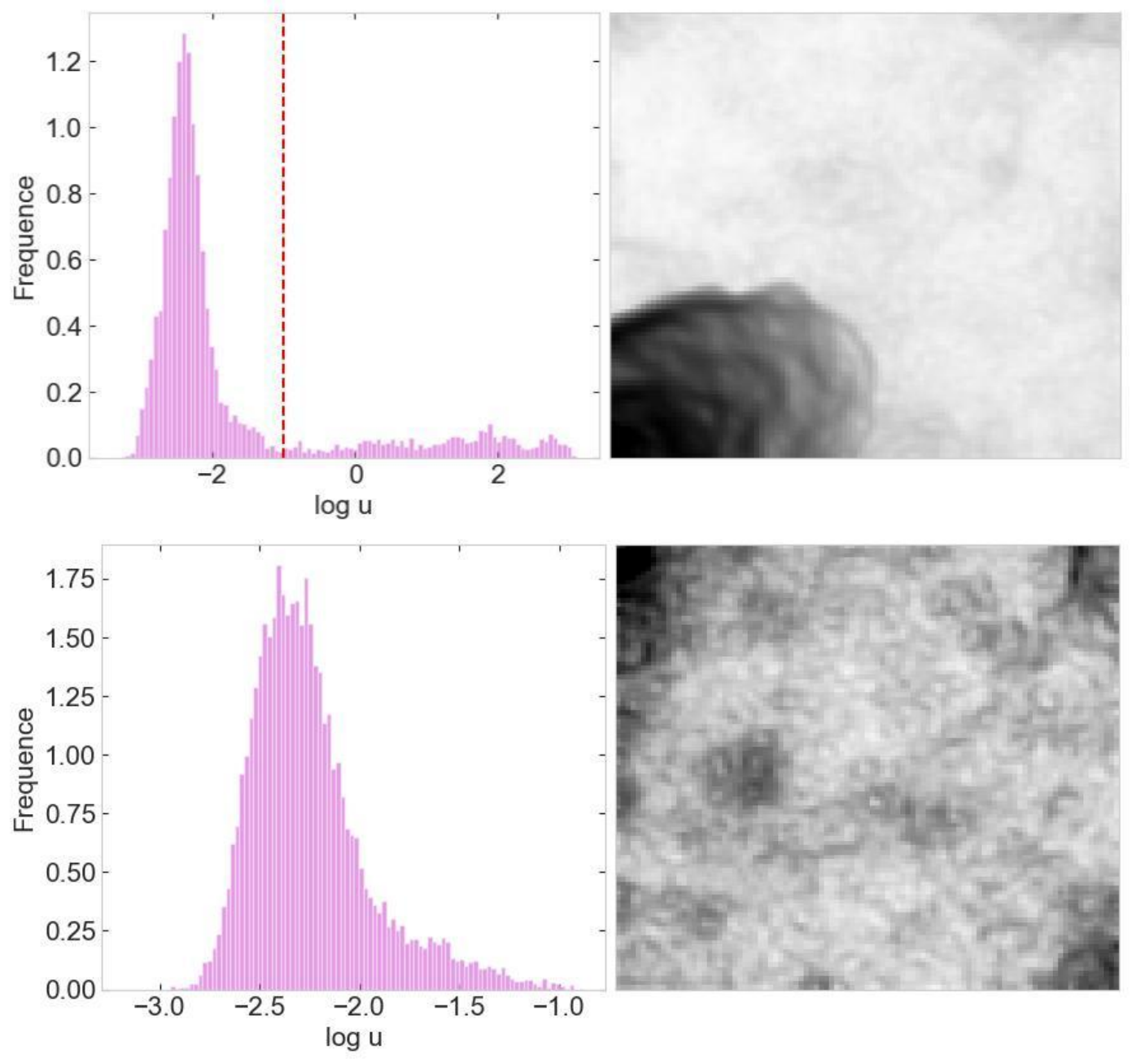}
    \caption{The distribution of logarithmic $\mathbf{u_{mtx}}$ elements and the 2-D neuron map obtained by StarGO in the normalized space containing $E, \,J_{\phi},\, J_{R},\, J_{z} $, [Fe/H], [Mg/Fe], and [Al/Fe]. Note that a darker color in the 2-D neuron map represents a higher value of $u$. The top panels display the training results for the parent dataset, where the red-dashed line indicates a threshold of $u_{82\%}$. The bottom panel displays the training results for the refined dataset}
    \label{figure1}
\end{figure}

\section{Method} \label{3}
We use the adaptive clustering algorithm, Stars' Galactic Origins \citep[StarGO;][]{Yuan2018StarGO} to identify substructures within the inner halo in 7-D space. This algorithm has been used to discover the LMS-1 structure \citep{Yuan2020als}, identify the Cetus stream \citep{Yuan2019revealing} and dynamically tagged groups (DTGs) in the nearby stellar halo \citep{Yuan2020dynamic}. The StarGO algorithm combines the self-organizing map (SOM) and a novel adaptive group identification methodology. In this work, we adopt a $100\,\times\,100$ 2-D neural map where each grid of the map contains a neuron. A neuron positioned at $(a, b)$ is assigned a weight vector, which has the same dimensionality as the input data. The initial weight vectors for each neuron is randomised and the differences in weight vectors between neighboring neurons are denoted by a $100\,\times\,100$ matrix, $\mathbf{u_{mtx}}$, whose elements are defined as follows \citep{Yuan2018StarGO}:
\begin{equation}
u_{a, b}=\left|\mathbf{w}_{a \pm 1, b}-\mathbf{w}_{a, b}\right|+\left|\mathbf{w}_{a, b \pm 1}-\mathbf{w}_{a, b}\right| 
\label{eq:1}
\end{equation}
Input data of a star in above neural network is considered as a vector $\mathbf{v}$. The best matching unit (BMU) on the map is defined with a neuron on the 2-D map with minimal value of $\mid \mathbf{w} - \mathbf{v} \mid$. The change in the weight vectors of each neuron $\mathbf{dw}$ is influenced by their distances to the BMU, $d_{a, b}$:
\begin{equation}
\mathbf{d} \mathbf{w}_{a, b}=\alpha_{q} \exp \left(-\frac{d_{a, b}{ }^{2}}{\sigma_{q}^{2}}\right)\left(\mathbf{v}-\mathbf{w}_{a, b}\right) 
\label{eq:2}
\end{equation}
where $d_{a, b}=\sqrt{\left(a-a_{i}\right)^{2}+\left(b-b_{i}\right)^{2}}$ with the BMU located at $(a_{i},\,b_{i})$. The learning rate and neighboring influence of neurons around the BMU for the $q$-th iteration are characterized by $\alpha_{q}$ and$\,\sigma_{q}$ respectively:
\begin{equation}
\alpha_{q}=\alpha_{0}\left(1-q / N_{\text {iter }}\right), \quad \sigma_{q}=\sigma_{0}\left(1-q / N_{\text {iter }}\right) 
\label{eq:3}
\end{equation}
In the process of iteration, the matrix $\mathbf{u_{mtx}}$ is guaranteed to converge when $\mathbf{dw}/\mathbf{w}\to 0$. We used an iteration number of $N_{iter} = 400$, as the matrix $\mathbf{u_{mtx}}$ converged after this point.
\par Neurons mapped to stars in highly-clustered regions tend to have similar characteristics, resulting in lower values for the $\mathbf{u_{mtx}}$ elements ($u$) and vice versa. The left panels in Figure \ref{figure1} show the distribution of $u$ and the right panels show the resulting 2-D neural map where grayer part represents higher values of $u$. After SOM algorithm, stars are grouped based on 2-D neuron map under threshold $u_{thr}$ (i.e., stars with $u\,<u_{thr}$ could be grouped). Stars belonging to substructures that are dispersed in the 7-dimensional space will be discarded during the process of decreasing $u_{thr}$. Consequently, some substructures may not be identified by this algorithm.

\par The 7-D space includes orbital energy ($E$), actions ($J_{\phi}$, $J_{R}$, $J_{z}$), where $J_{\phi}$ equals to $(L_{z})$ in ${\tt  MWPotential2014}$ potential model \citep{Bovy2015}, as well as we incorporate chemical abundances of [Fe/H], [Mg/Fe] and [Al/Fe] to further distinguish different stellar groups and their in-situ/ accreted origins \citep{Hawkins2015using}. Specifically, [Mg/Fe] and [Al/Fe] have been used to discuss the origin of structures \citep{Hasselquist2021}. It is important to note that although different accreted substructures can be identified through clustering in the above kinematic-related spaces, distinguishing the progenitors of these substructures through clustering results can be misleading, as interpreting different clumps in such spaces as originating from distinct satellite galaxies is not physically motivated \citep[e.g.,][]{Jean-Baptiste2017onJ,Amarante2022ApJ,Pagnini2023A&A,Khoperskov2023A&A,Mori2024eas}. Specifically, a single satellite can distribute its contents over a wide region of kinematic spaces due to dynamical friction. This process becomes significant when the satellite's mass is sufficiently large relative to the MW, leading to the breakdown of \( E - L_z \) conservation. As a result, spatially separated substructures in kinematic spaces may share a common origin from the same satellite.
\par According to \citet{Yuan2022the}, we reshuffled the training samples within each dimension of the input space to generate a smooth halo sample. As a result, this halo sample exhibits the same distribution as the original data in the input space, while eliminating any correlations between the input dimensions arising from substructures. Subsequently, for each identified group, their contamination is evaluated and significance is calculated \citep{Yuan2022the}. We consider a group to be valid if its significance exceeds 5$\sigma$. Previous studies \citep{Naidu2020evidence,Malhan2022the,Horta2023the,Ye2024dynamical} have developed relatively comprehensive criteria for identifying substructures. The criteria employed in this study are presented in Table \ref{table1}, where Galactocentric cylindrical velocities ($V_{z}$, $V_{\phi}$ and $V_{r}$ are Galactocentric cylindrical vertical, rotational and radial velocities, respectively), angle $\left| \beta_{GC} \right|$ to the Sagittarius dwarf spheroidal (dSph) galaxy, left-handed heliocentric Cartesian coordinates ($X, Y, Z$), Cartesian coordinates centered on the Sgr dSph plane ($X_{Sgr}, Y_{Sgr}, Z_{Sgr}$), eccentricity ($e$), total action ($J_{tot}$), maximum height above the Galactic plane ($Z_{\mathrm{max}}$), and chemical abundance [X/H] ([X/Fe]) are included. It is important to reiterate that the criteria in Table \ref{table1} primarily focus on identifying regions in kinematic spaces as substructures, which does not mean that these substructures are independent or have different progenitors \citep[e.g.,][]{Jean-Baptiste2017onJ,Amarante2022ApJ,Pagnini2023A&A,Khoperskov2023A&A,Mori2024eas}. Associating different substructures is not the main focus of this paper, as our goal is to identify potentially new substructures. Furthermore, those substructures associated with known progenitors, based on their characteristics in kinematic spaces, should be regarded as tentative memberships. When considering kinematic features, correlations between substructures should also be treated as tentative relationships.

\begin{table}
	
	\footnotesize
	\setlength{\tabcolsep}{4pt}
	\renewcommand{\arraystretch}{1.5}
	\centering
	\begin{tabular}{lcc}
		\hline
		Name & Selection Criteria & Criteria Reference \\
		\hline
		Sagittarius Stream   & $1.8\,< L_{z,Sgr}\,<\,14 \,(\times10^{3}\, \mathrm{kpc}\,\mathrm{\,km\,s^{-1}}),\,-150\,< V_{z,Sgr}\,<\,80 \,(\,\mathrm{km\,s^{-1}}),\, $ &\citet{Horta2023the}\\&$ X_\mathrm{Sgr}\,>\,0 \,(\,\mathrm{kpc}) \,$or$\,X_\mathrm{Sgr}\,<\,-15 \,(\,\mathrm{kpc}),\,Y_\mathrm{Sgr}\,>\,-5 \,(\,\mathrm{kpc}) \,$or$\,Y_\mathrm{Sgr}\,<\,-20 \,(\,\mathrm{kpc}), $\\& $Z_\mathrm{Sgr}\,>\,-10 \,(\,\mathrm{kpc}), \,pm_{\alpha}\,>\,-4 (mas),\,d_{\odot}\,>\,10 \,(\,\mathrm{kpc}),\, \left| \beta_{GC} \right| < 30^{\circ}$\\ \hline
		Gaia-Sausage-Enceladus& $e >\,0.7 , \,(J_{z}-J_{R})/J_{tot}\,<\,-0.5$ & \citet{Ye2024dynamical}\\ \hline
		Sequoia+Arjuna+I'itoi&$ E >\,-3\times10^{4} \,\mathrm{km^{2}s^{-2}}
		,\,L_{z}\,<\,-1.4\times10^{3}\,\mathrm{kpc}\,\mathrm{km\,s^{-1}}$ &\citet{Ye2024dynamical}\\ \hline
		Thamnos& $-7\,<\,E <\,-3 (\times10^{4} \,\,\mathrm{km^{2}s^{-2}}),\,L_{z}\,< 0 \,\mathrm{kpc}\,\mathrm{km\,s^{-1}} 
		, \,e < 0.7 $&\citet{Horta2023the}\\ \hline
		Helmi Streams&$0.75\,<L_{z}\,<\,1.7 \,(\times10^{3}\,\mathrm{kpc}\,\mathrm{km\,s^{-1}}),\,1.6\,<L_{\perp}\,<\,3.2\, (\times10^{3}\,\mathrm{kpc}\,\mathrm{km\,s^{-1}})$&\citet{Horta2023the}\\ \hline
		Aleph& $175 < V_{\phi} <\,300 \,(\,\mathrm{km\,s^{-1}})
		,\, \left| V_{r}\right|  <\,75\, (\,\mathrm{km\,s^{-1}})
		,\, $[Fe/H]$>$$-$0.8 dex,&\citet{Naidu2020evidence}\\

		&\, $J_{z}-J_{R}\,>\,0$,\, [Mg/Fe]$<$0.27 dex \\  \hline
		Wukong/LMS-1&$ J_{\phi}/J_{tot}\,> 0 
		,\,\,(J_{z}-J_{R})/J_{tot}\,>\,-0.5$$\,\,\,, 0.4\,< e < 0.7 $, [Fe/H] $<-$1.45 dex &\citet{Horta2023the};\\
        &&\citet{Ye2024dynamical}\\
        \hline
		Nyx&$110 < V_{r} <\,205\, (\,\mathrm{km\,s^{-1}})
		, \,90 < V_{\phi} < 195 \,(\,\mathrm{km\,s^{-1}}), $&\citet{Horta2023the}\\&$
		 \left| X \right|  <3 \,\mathrm{kpc},\left| Y \right|  <\,2 \,\mathrm{kpc} ,\left| Z \right|  <\,2 \,\mathrm{kpc}$\\ \hline
		Icarus&[Fe/H]$<\,-$1.05 dex,\,
		[Mg/Fe]$<$0.2 dex, $L_{\perp}\,<\,450 \,\mathrm{kpc}\,\mathrm{km\,s^{-1}}$
		,&\citet{Horta2023the}\\&
		$1.54\,< L_{z}\,<\,2.21 \,(\times10^{3}\,\mathrm{kpc}\,\mathrm{km\,s^{-1}}),\,
		 e< 0.2,\,
		Z_{max}\,<\,1.5 \,\mathrm{kpc}$\\ \hline
		Pontus&$-470\,<L_{z}\,< 5 \,(\,\mathrm{kpc}\,\mathrm{km\,s^{-1}} )
		,\, 245\,<J_{R}\,<\,725\, (\,\mathrm{kpc}\,\mathrm{km\,s^{-1}} )$&\citet{Horta2023the}\\&
		$,\, 115\,<J_{z}\,<\,545 \,(\,\mathrm{kpc}\,\mathrm{km\,s^{-1}} )
		,\, 390\,<L_{\perp}\,<\,865 \,(\,\mathrm{kpc}\,\mathrm{km\,s^{-1}})$\\&
		,\, $ 0.5\,<e< 0.8 , $ $E\,<\,-4\,(\times10^{4} \,\,\mathrm{km^{2}s^{-2}}),$
		\,[Fe/H]$<\,-$1.3 dex\\ \hline
		Cetus& $0.4\,<e< 0.6\, 
		,\ 605\,<J_{R}\,<\,1075 \,(\,\mathrm{kpc}\,\mathrm{km\,s^{-1}})$ 
		,&\citet{Malhan2022the}\\& $1360\,<L_{z}\,<\,2700\, (\,\mathrm{kpc}\,\mathrm{km\,s^{-1}})
		, \,1835\,<J_{z}\,<\,2820\, (\,\mathrm{kpc}\,\mathrm{km\,s^{-1}}) $
		,\\& $2905\,<L_{\perp}\,<\,4635 \,(\,\mathrm{kpc}\,\mathrm{km\,s^{-1}}) $ \\
            \hline
            Metal-Weak\,Thick\,Disk&$-$2.5$<$[Fe/H]$<\,-$0.8 (dex),\,0.25$<$[Mg/Fe]$<$0.45 (dex),\, $J_{\phi}/J_{tot}\,>\,0.5$&\citet{Naidu2020evidence}\\
            \hline
            
	\end{tabular}
        \caption{The table presently exhibits the substructures discovered thus far, alongside the stringent criteria employed for their identification. It is imperative to note that all orbital parameter values sourced for this purpose have been meticulously retrieved from the APOGEE-astroNN catalog.}
        \label{table1}
\end{table}

\section{Results and Analysis} \label{4}
\par 
Based on the selection criteria above, we primarily excluded possible disk stars by limiting [Fe/H] $<\,-0.8$ dex \citep{Hawkins2015using}. After the StarGO algorithm, we obtained the distribution of $u$ and 2-D neuron map exhibited in the top panels in Figure \ref{figure1}. A specific data segment in the bottom-left corner of the top row’s 2-D map, characterized by high $u$ values, negatively impacted substructure identification. To address this, we applied a threshold of $u_{82\%}$ (marked by the red line on the top-left panel) to exclude this segment. We further remove possible disk stars in the clustering results that overlap with Galactic disks in the chemical planes \citep{Hawkins2015using}, leaving a refined sample of 6,936 stars. The bottom panels of Figure \ref{figure1} shows the distribution of $u$ and 2-D neuron map obtained from StarGO.

\par
\begin{figure}
    \centering
    \includegraphics[width=1\linewidth]{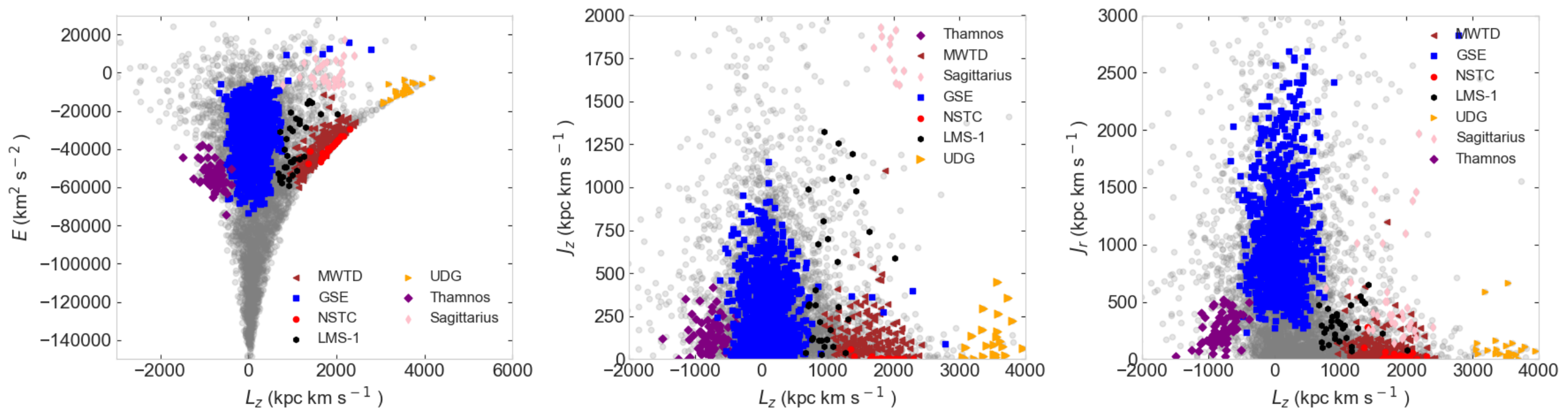}
    \caption{The distribution of different substructures in the $(\mathit{\textbf{J} , E)}$ space.}
    \label{figure2}
\end{figure}

\begin{figure}
    \centering
    \includegraphics[width=0.7\linewidth]{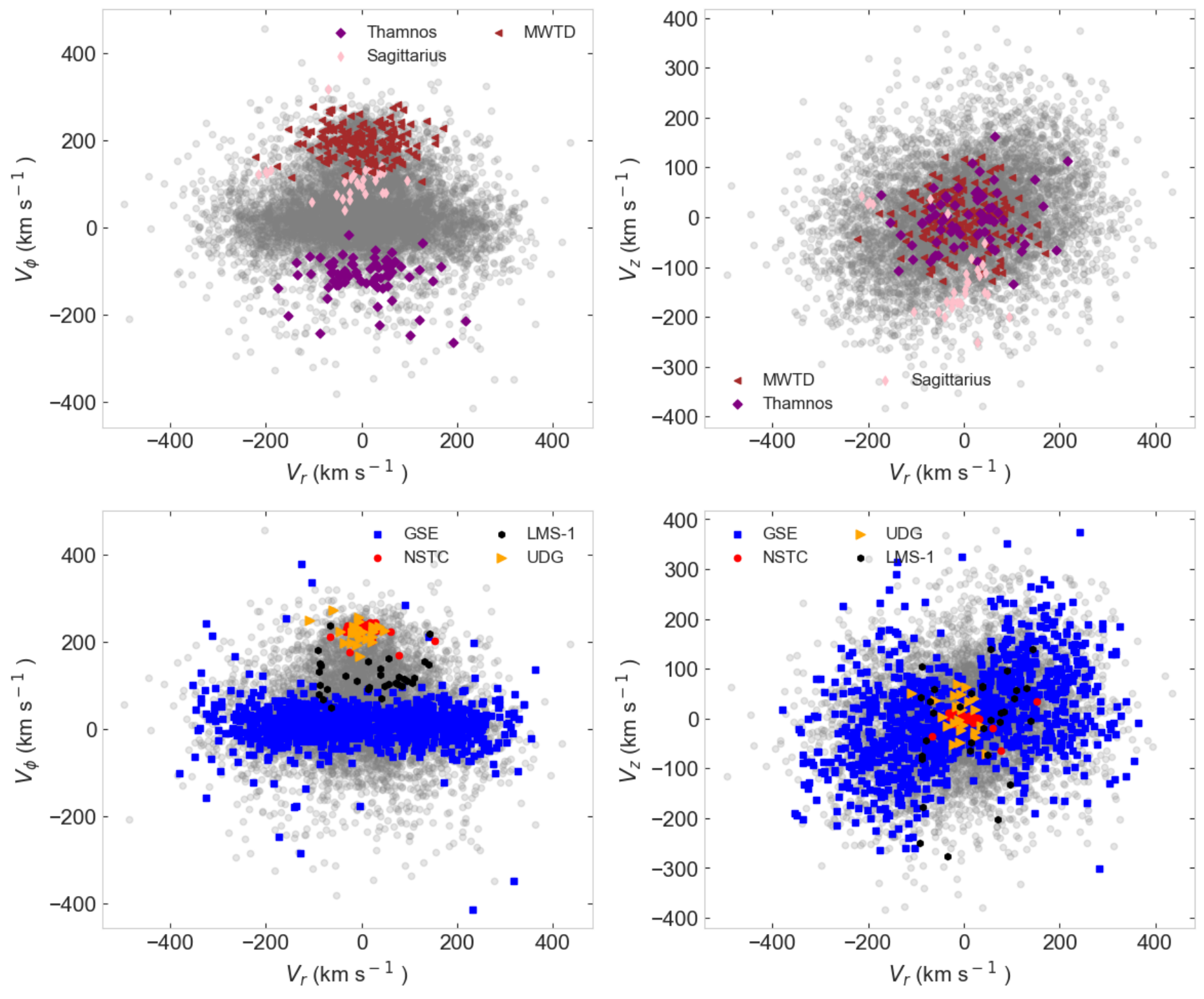}
    \caption{The distribution of substructures in the velocity space. Left column: substructures in $(V_{r},V_{\phi})$ space. Right column: substructures in $(V_{r},V_{z})$ space.  Gray dots come from the refined parent sample.}
    \label{figure3}
\end{figure}

\begin{figure}
    \centering
    \includegraphics[width=0.9\linewidth]{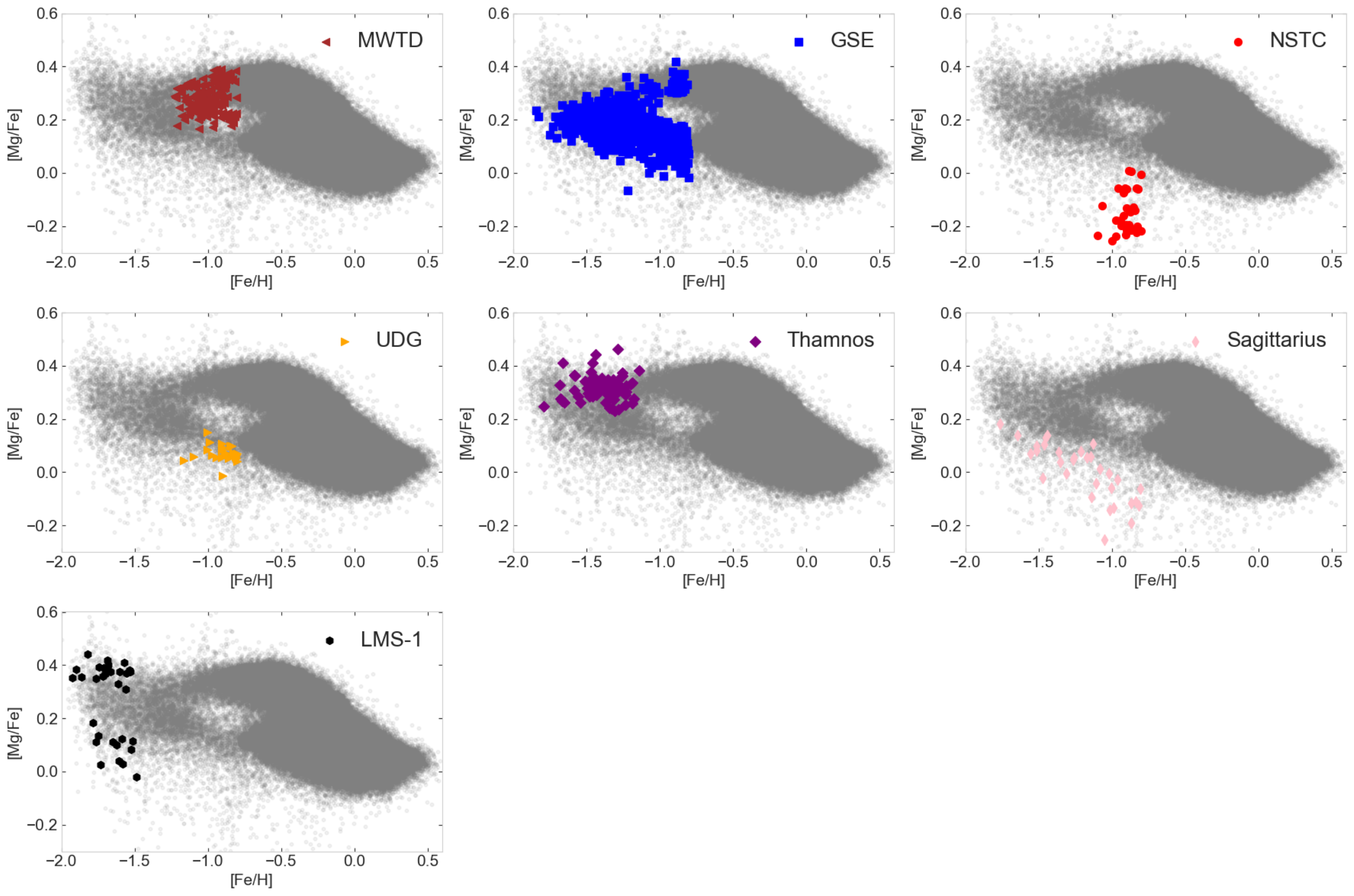}
    \caption{The distribution of identified substructures in the [Fe/H]-[Mg/Fe] plane.}
    \label{figure4}
\end{figure}

\begin{figure}
    \centering
    \includegraphics[width=0.9\linewidth]{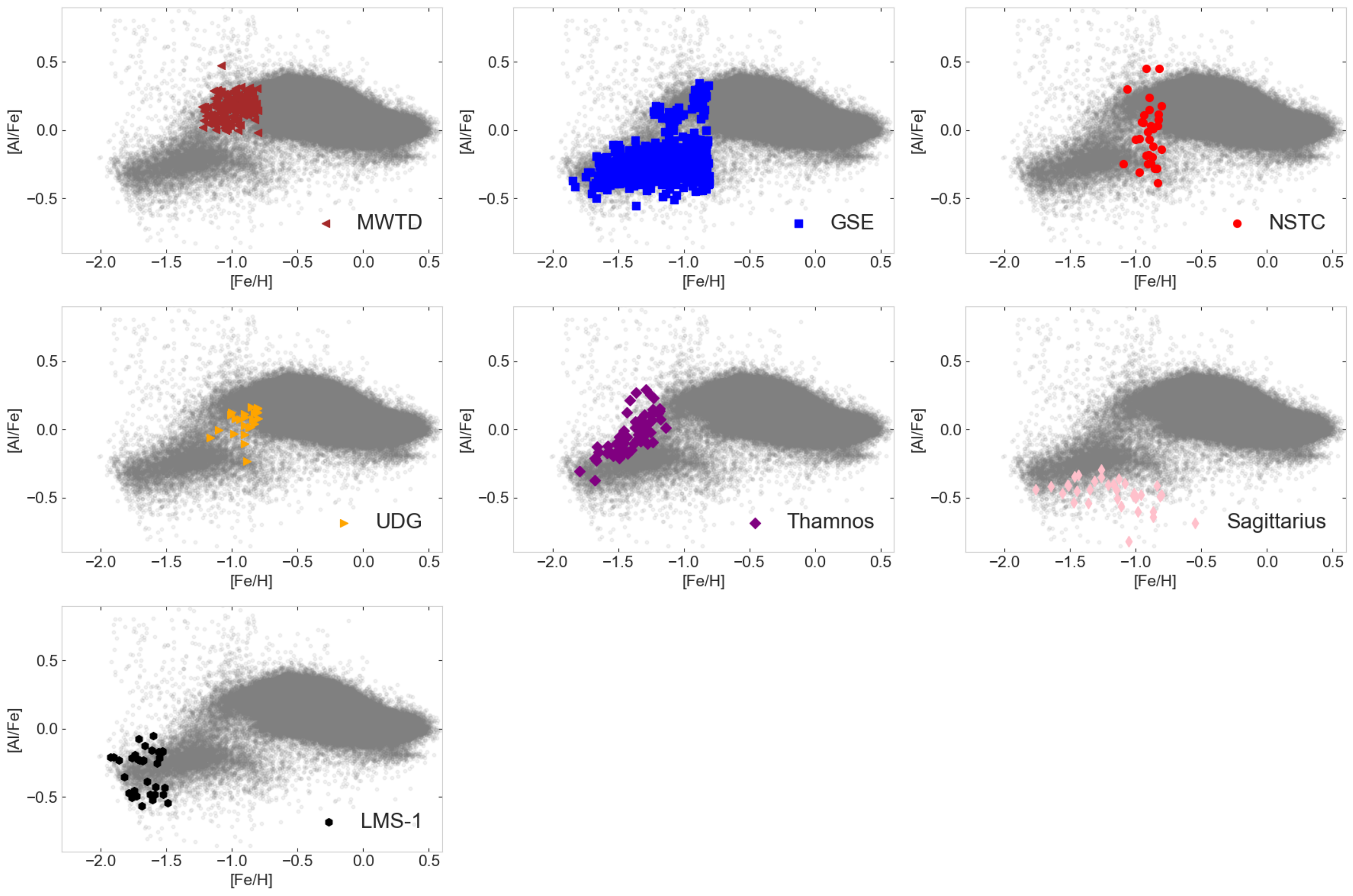}
    \caption{The distribution of identified substructures in the [Fe/H]-[Al/Fe] plane.}
    \label{figure5}
\end{figure}

 Using the refined sample and selection criteria, we identified 42 groups based on the clustering results. These groups include five well-known substructures. For clarity, the name of NSTC refers to the possible new substructure candidates, while UDG denotes undefined group related to known substructures, as discussed in Subsection \ref{4.1} and Subsection \ref{4.2}. The kinematic and chemical distributions of these substructures are visualized in Figure \ref{figure2}$-$\ref{figure5}, spanning $(\textbf{J},E)$ space, velocity space, and the chemical space of ([Fe/H], [Mg/Fe], [Al/Fe]). For clarity, only substructures identified by valid groups are presented. Notably, Mg abundance serves as our $\alpha$-element tracer. Regarding the abbreviations used: Sgr denotes Sagittarius Stream, NSTC stands for new substructure candidate, GSE signifies Gaia-Sausage-Enceladus, MWTD is Metal-Weak Thick Disk and UDG indicates undefined group.


\subsection{Substructures From the Refined Sample}\label{4.1}

$\bullet $ Thamnos: distinct from Sequoia \citep{Barb2019as,Matsuno2019origin,Myeong2019evidence}, it is theorized to be a remnant of an accreted dwarf galaxy based on its strong retrograde rotation, high binding energy and low metallicity \citep{Koppelman2019Multiple}. Simulations suggest its progenitor had a stellar mass of approximately $10^{6}\, \mathrm{M}_{\odot}$. With such a low mass, Thamnos's lower orbital energy compared to GSE indicates a very early accretion event \citep{Koppelman2019Multiple, Naidu2020evidence}. In line with the approach adopted by \citet{Horta2023the}, we treat Thamnos as an individual substructure and identify its member stars according to the following criteria: $-7\,<\,E <\,-3 (\times10^{4} \,\,\mathrm{km^{2}s^{-2}}),\, L_{z}\,< 0 \,\mathrm{kpc}\,\mathrm{km\,s^{-1}},\, e < 0.7 $. The selection criteria based on $E$ identifies stars with intermediate orbital energies, differing from those used in \citep{Horta2023the} due to the application of a different Galactic potential. Our work identified a group of 70 stars representing Thamnos, as presented in Figure \ref{figure2}$-$\ref{figure5}.
\par 
$\bullet $ Metal-Weak Thick Disk: the Galactic thick disk can be divided into a canonical thick disk and Metal-Weak Thick Disk (MWTD)  \citep{Chiba_2000}. By analyzing the kinematic and chemical properties of a stellar sample with Galactocentric distance range in 7 kpc $< R_{\mathrm{gal}} <$ 10 kpc, \citet{Carollo2019evidence} propose three scenarios for the origin of the MWTD: one scenario suggests MWTD stars originate from thick disk stars heated by a GSE-like merger, another possibility is that MWTD stars migrated from smaller radii due to bar instability or spiral-arm formation; the third scenario posits that the MWTD is debris from a massive accreted satellite, circularized by dynamical friction. In this study, we identified MWTD member stars according to the criteria established in previous literature \citep{Naidu2020evidence}, specifically: $-$2.5 $<$[Fe/H]$<$ $-$0.8 (dex), 0.25 $<$[Mg/Fe]$<$ 0.45 (dex), and $J_{\phi}/J_{tot}\,>\,0.5$. Consequently, we have obtained 70 MWTD member stars. As shown in the first panel of Figure \ref{figure4}, the MWTD is positioned adjacent to the high-$\alpha$ disk, sharing a similar [Mg/Fe] range with the thick disk, in agreement with \citet{Naidu2020evidence}.

$\bullet $ Gaia-Sausage-Enceladus: it was initially discovered by \citet{Belokurov2018Co} and \citet{Helmi2018the}. \citet{Belokurov2018Co} suggested that the highly radial orbits of GSE, with most stars having an eccentricity $e > 0.7$, could result from a dramatic radialization of the orbit of its massive progenitor, a process further amplified by the action of the growing disk. Subsequent studies \citep{Helmi2018the, Koppelman_2018, Myeong2018the, Haywood2018in, Mackereth2019the} have further indicated that GSE may represent the last major accretion event in the formation of the MW. It is named for its blob-shaped/sausage-shaped distribution near $V_{\phi}\,\sim\,0$ in the $V_{r} -V_{\phi}$ space \citep{Belokurov2018Co,Belokurov2020MNRAS}. \citet{Wu2022contribution} found that this substructure dominates the inner halo, comprising 41\% - 71\% of its stars. In this work, we selected GSE members using $e > 0.7 $ and $ (J_{z}-J_{R})/J_{tot}\,<\,-0.5$ \citep{Ye2024dynamical} to exclude Pontus \citep{Malhan2022the}. We identified 20 groups of GSE stars (1058 stars) with 18 of them (1035 stars) exceeding $5\sigma$ significance. High-significance GSE members are presented in Figure \ref{figure2} to Figure \ref{figure5}. The $L_{z}$ distribution shown in Figure \ref{figure2} is in good agreement with the selection criteria of $|L_{z}| < 500\,\mathrm{kpc\,km\,s^{-1}}$ in \citet{Horta2023the}, and the velocity distribution is displayed in the bottom-left panel in Figure \ref{figure3}.

\par 
As the clustering threshold decreased, the relations between stars tightened, causing large groups, such as UDG-7 (initially containing 468 stars at a threshold of $u_{thr}\,=\,u_{39\%}$), to fragment into smaller ones: MWTD-4, MWTD-5, MWTD-6, MWTD-7, GSE-18 and GSE-19 (see Table \ref{tab:idectification1}). The tentative memberships assigned in this paper for substructures are therefore highly dependent on the threshold used, and the sample of member stars may also be incomplete. Additionally, unlike the substructures such as Thamnos, GSE member stars are identified under several thresholds. Some member stars of GSE are even separated from groups along with other substructures (specifically, Thamnos and MWTD). This suggests significant correlations in the kinematic and chemical characteristics of member stars among these substructures. In addition, it is possible that the GSE could be a composite substructure resulting from multiple accretion events, as suggested by \citet{Donlon2022ASO}.

\begin{table}[H]
        \caption{\enspace Groups of Different Thresholds} 
	\footnotesize
	\setlength{\tabcolsep}{2pt}
	\renewcommand{\arraystretch}{1.5} 
        \centering
	\begin{tabular}{|lc|cccc|c|c|c|}
		\hline
		$u_{thr}$&& Group Name &&&& Number& Contamination& Significance  \\
		\hline
		$u_{39\%}$&&UDG-7&&&&468&36.97\%&$>5\sigma$\\
		\hline
		&$u_{31\%}$&&MWTD-4&&&16&37.5\%&99.968\%\\
		\hline
            &$u_{20\%}$&&GSE-18&&&22&27.27\%&$>5\sigma$\\
            &&&MWTD-5&&&16&37.5\%&99.968\%\\
		&&&MWTD-6&&&69&34.78\%&$>5\sigma$\\
            &&&MWTD-7&&&145&22.06\%&$>5\sigma$\\
            &&&GSE-19&&&13&7.69\%&$>5\sigma$\\
  
		\hline
		
		\hline
        
	\end{tabular}
        \\
        \parbox{\textwidth}{
        \vspace{6mm}
        $u_{thr}$ in this table represents threshold. The third column represents the number of member stars of groups. The last two columns represents the expected contamination from the smooth halo sample and significance of the groups respectively. The thresholds and group names of smaller groups separated from larger one are listed on the right side of each column.}

	\label{tab:idectification1}
\end{table}

\par
$\bullet $ Sagittarius Stream: it is an early-discovered substructure by \citet{Ibata1994ads} and has been extensively characterized in subsequent studies \citep{Ibata2001galactic,Majewski2003atm,Belokurov2006the,Carlin2018chemical,Antoja2020an,Ibata2020apl,Vasiliev2020the}. It exhibits high energy and prograde orbits, with heliocentric distances concentrated around 23 kpc \citep{Vasiliev2020the}. Additionally, the Sgr shows relatively low $\alpha$ abundance, even compared to the low-$\alpha$ disk. \citet{Horta2023the} and \citet{Hasselquist2021} have also shown that there is an upside-down $\alpha$-``knee'' at [Fe/H] $\sim -0.8$ dex in the [Fe/H]-[Mg/Fe] plane, likely caused by a burst in star formation due to its interaction with the MW. According to \citet{Horta2023the}, the criteria employed in this work are as follows:
$\left| \beta_{GC} \right| < 30^{\circ}$, $1.8\,< L_{z,Sgr} < 14\, (\times10^{3}\,\mathrm{kpc}\,\mathrm{km\,s^{-1}})$,  $-150\,< V_{z,Sgr}\,<\,80 \,(\,\mathrm{km\,s^{-1}})$, $X_\mathrm{Sgr}\,>\,0 \,(\,\mathrm{kpc}) \,$or$\, X_\mathrm{Sgr}\,<\,-15  \,(\,\mathrm{kpc})$, $Y_\mathrm{Sgr}\,>\,-5 \,(\,\mathrm{kpc}) \,$or$\,Y_\mathrm{Sgr}\,<\,-20  \,(\,\mathrm{kpc})$, $Z_\mathrm{Sgr}\,>\,-10 \,(\,\mathrm{kpc})$, $pm_{\alpha}\,>\,-4 (mas)$, and $d_{\odot}\,>\,10 \,(\,\mathrm{kpc})$. In this study, we identified 32 member stars belonging to Sagittarius Stream (refer to Figure \ref{figure2} $-$\ref{figure5}). This relatively small number is due to the scarcity of the Sagittarius Stream in the solar neighborhood. Additionally, due to the imposed limit on [Fe/H], the characteristic $\alpha$-``knee'' feature in \citet{Horta2023the} is absent in our study.
\par
$\bullet $ LMS-1/Wukong: this low-mass stream was first discovered by \citet{Yuan2020als} and named LMS-1, and later independently discovered by \citet{Naidu2020evidence}, who referred to it as Wukong. It is characterized by a relatively wide eccentric distribution and a very metal-poor nature with [Fe/H] $< -1.45$ dex. To avoid selecting GSE member stars, \citet{Horta2023the} added a criterion of $e < 0.7$, along with a criterion of $e > 0.4$ to distinguish it from disk stars. \citet{Naidu2020evidence} suggests that LMS-1 may have been accreted by the Milky Way along with three metal-poor globular clusters: NGC 5024, NGC 5053, and ESO 280-SC06. In this work, we adopt the selection according to \citet{Horta2023the} and \citet{Ye2024dynamical} but without the cut on vertical height above the plane ($|z\,>|$ 3 kpc), as follows: $ J_{\phi}/J_{tot}\,> 0 ,\,(J_{z}-J_{R})/J_{tot}\,>\,-0.5$, $\, 0.4\,< e < 0.7 $ and [Fe/H] $<-$1.45 dex. Applying these criteria, we identified two distinct groups of stars within the LMS-1, comprising 32 members. The distribution of these stars are very consistent with those in \citet{Naidu2020evidence} as shown in Figure \ref{figure2} $-$ \ref{figure5}.
\par $\bullet $ UDG: containing 24 member stars, this group has a significance greater than $5\sigma$. Its kinematic and chemical distribution are as follows:
$E\,\sim\,[-2.0,\,0]$$\times 10^{4}$ km$^2$ s$^-$$^2$ , $L_{z}\,\sim\,[3000,\,4000]\,\mathrm{kpc}\,\mathrm{km\,s^{-1}}$, $J_{r}\,\sim\,[0,\,650]\,\mathrm{kpc}\,\mathrm{km\,s^{-1}}$, $J_{z}\,\sim\,[0,\,500]\,\mathrm{kpc}\,\mathrm{km\,s^{-1}}$, [Fe/H]$\,\sim\,$[$-$1.2, $-$0.8] dex, [Mg/Fe]$\,\sim\,$[$-$0, 0.16] dex, [Al/Fe]$\,\sim\,$[$-$0.2, 0.2] dex. UDG shows disk-like dynamics (see Figures \ref{figure2} and \ref{figure3}) and highly circular orbits with most eccentricities $<$ 0.3. Its distribution in the [Fe/H]-[Mg/Fe] plane is located next to the low-$\alpha$ disk and the GSE. Given its shared kinematic and chemical abundance characteristics ([Mg/Fe], [Al/Fe]) with Aleph \citep{Naidu2020evidence}, we initially considered this substructure as a potential candidate for Aleph. Notably, this substructure exhibits a relatively significant vertical action ($J_z$), with most of its members having $J_z > J_R$, which supports its classification as Aleph. However, the metal-poor nature of UDG's members ([Fe/H] $< -$0.8 dex) suggests that they may represent outlier member stars of Aleph in chemical space.

\par $\bullet$ NSTC: it exhibits significance levels exceeding 5$\sigma$ and is distinguished by its unique low-$\alpha$ and disk-like dynamical properties. \citet{Horta2023the} define in-situ stars as those with [Fe/H] $>-0.8$ dex and [Mg/Fe] $<0.22$ dex, or [Fe/H] $>-1.05$ dex and [Mg/Fe] $>0.22$ dex. With this criterion, NSTC would not be classified as in-situ. Notably, NSTC resides near the upper limit of [Fe/H] ($\sim\,-$0.8 dex), suggesting the possibility that it may contain metal-rich ([Fe/H] $> -$0.8 dex) member stars. To figure out the properties of NSTC, we discarded the [Fe/H] constraint with a dataset containing 222,388 stars. Subsequently, we isolated and removed a group with a large number of stars that significantly overlaps with the Galactic disk in the chemical planes, based on the clustering process by StarGO, refining our sample to 10,294 stars. As clustered in the 7-D space, these removed stars likely represent disk stars that are unnecessary in this work. With the newly-refined dataset, we reapplied the above clustering procedure to identify substructures. In addition to the above known substructures, several other prominent groups/substructures are summarized in Table \ref{table2}.

\begin{table} [H]
        \centering
	\caption{\enspace New Groups at Different Thresholds Based on the Newly-Refined Dataset}
	\footnotesize
	\setlength{\tabcolsep}{2pt}
	\renewcommand{\arraystretch}{1.5}
	\centering
	\begin{tabular}{|lc|cccc|c|c|c|}
		\hline
		$u_{thr}$&&Group Name&&&&Number&Contamination &Significance  \\
		\hline
		$u_{90\%}$\,\,\,\,&\,\,\,\,&UDG-15&&&&164&31.09\%&$>5\sigma$\\
		\hline
		&$u_{88\%}$&&NSTC-1&&&37&27.11\%&$>5\sigma$\\
		&&&NSTC-2&&&32&31.35\%&$>5\sigma$\\
            &&&LAS&&&80&17.5\%&$>5\sigma$\\
            \hline

           $u_{87\%}$\,\,\,\,&\,\,\,\,&NSTC-3&&&&40&27.49\%&$>5\sigma$\\
		\hline
           $u_{82\%}$&&Aleph&&&&71&23.94\%&$>5\sigma$\\
            \hline
           $u_{50\%}$&&HAS&&&&102&23.60\%&$>5\sigma$\\
            \hline
	\end{tabular}
	\label{table2}
\end{table}

\subsection{The Second Clustering Result}\label{4.2}
\begin{figure}
    \centering
    \includegraphics[width=1\linewidth]{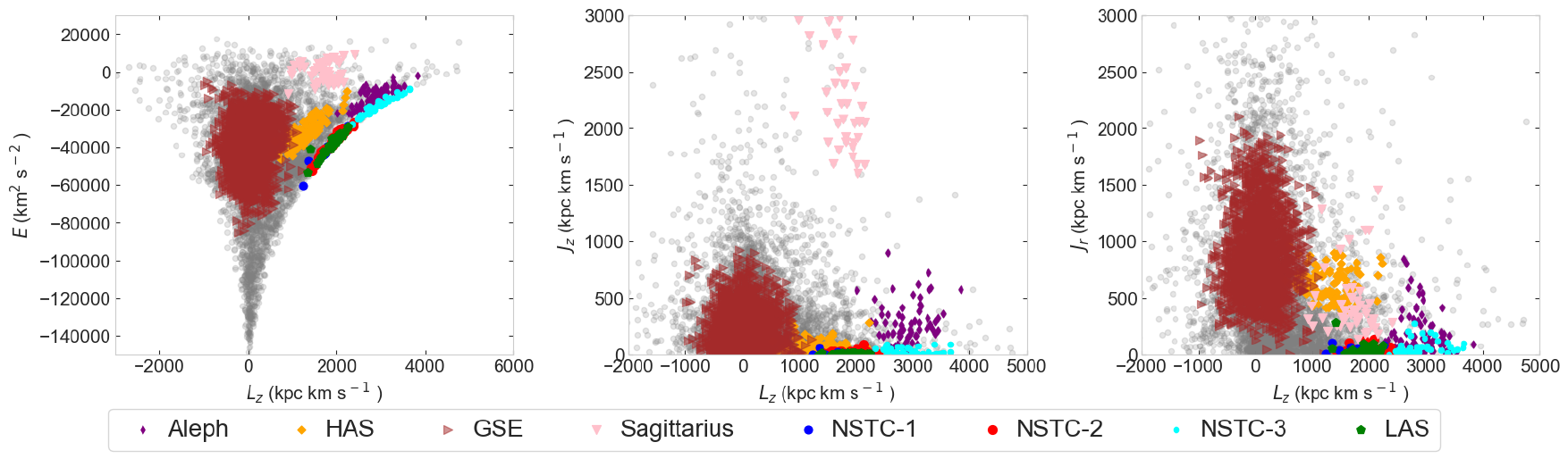}
    \caption{The distribution of GSE, Sgr and groups/substructures given in Table \ref{table2} in $(\mathbf{J}, E)$ space.}
    \label{figure6}
\end{figure}

\begin{figure}
    \centering
    \includegraphics[width=0.9\linewidth]{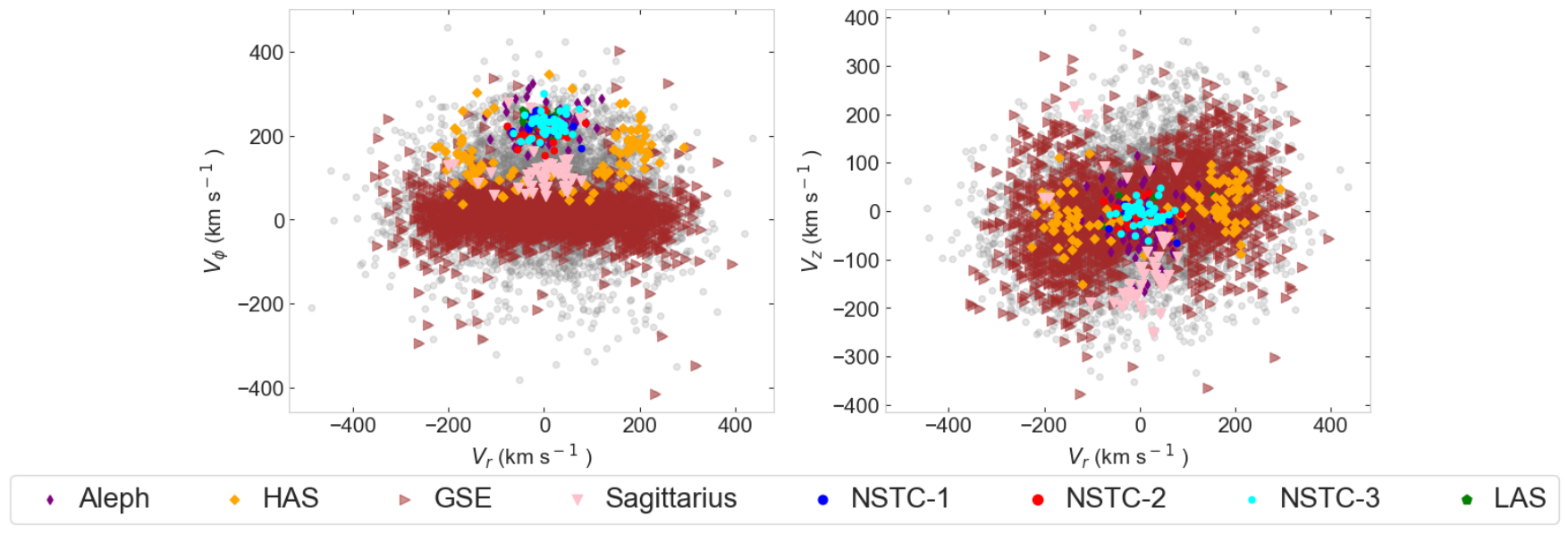}
    \caption{The distribution of GSE, Sgr and groups/substructures given in Table \ref{table2} in velocity space}. Gray dots come from the refined parent sample.
    \label{figure6.5}
\end{figure}

\begin{figure}
    \centering
    \includegraphics[width=1\linewidth]{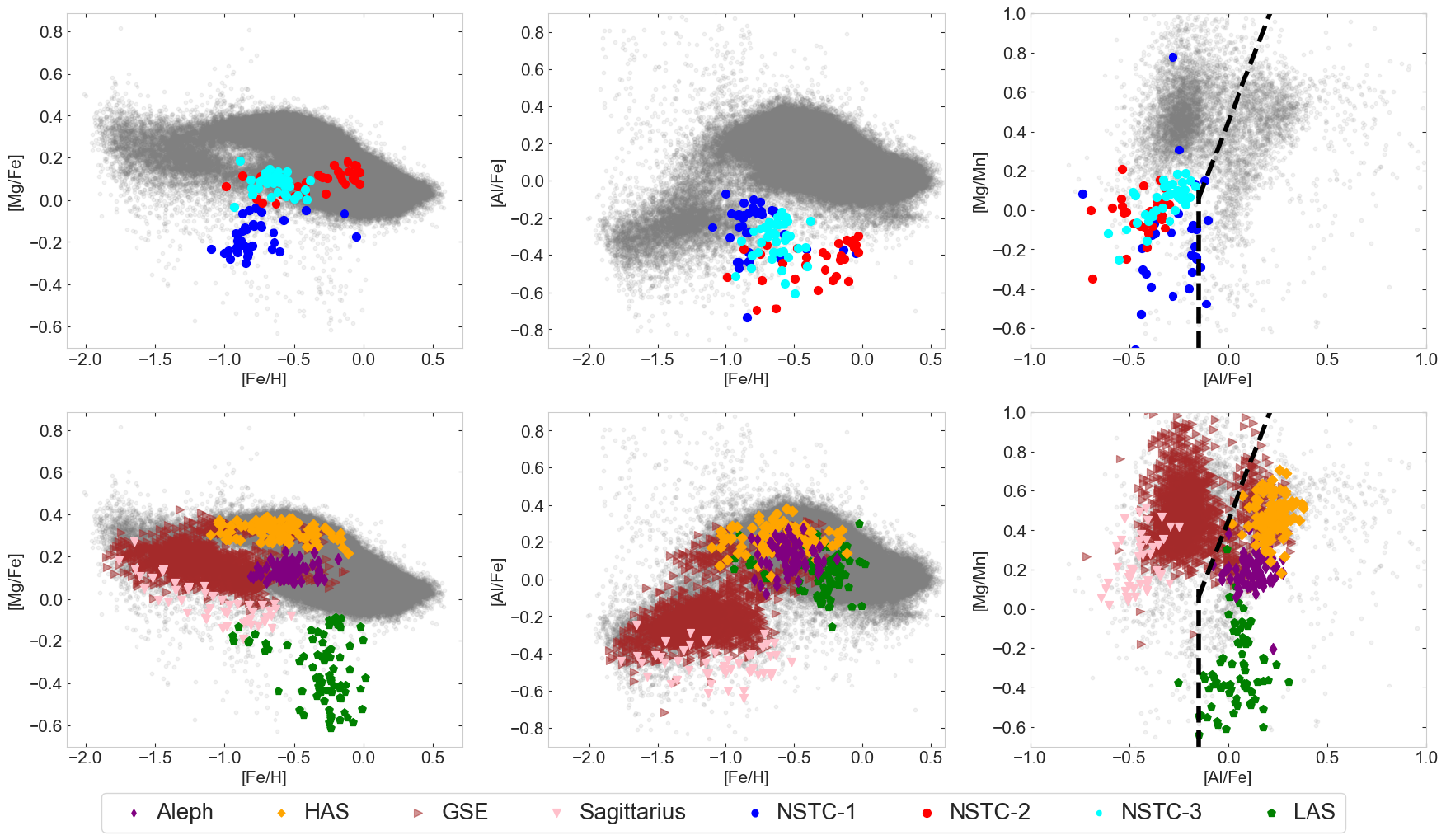}
    \caption{The element abundance distribution of GSE, Sgr and groups/substructures given in Table \ref{table2}. The NSTCs are displayed separately from other substructures to better exhibit its distribution. The black-dashed line in the right panels adopts the selection criteria outlined by \citet{das2020} and a [Al/Fe] threshold of $-$0.15 dex which is served as a demarcation between low and high [Al/Fe] abundance. Given the low [Al/Fe] can be associated with accretion events, the black-dashed line here could indicates the in-situ or accreted origin. Specifically, the left-hand side of this line designates the accreted component, whereas the in-situ component lies on the opposing side. Gray dots in the [Al/Fe]-[Mg/Mn] planes come from the newly-refined parent sample.}
    \label{figure7}
\end{figure}

 In Table \ref{table2}, LAS and HAS denote low-$\alpha$ abundance structure and high-$\alpha$ abundance structure respectively. Notably, NSTC-1 matches the previously identified NSTC with 13 common member stars. This is acceptable given the possible shifts in stellar correlations resulting from changes in data criteria, while still preserving the overall distribution characteristics of the substructures. Consequently, we suggest that NSTC-1 and NSTC represent the same substructure with NSTC-1 approaching a more general distribution characteristic without metallicity constraints. As shown in Table \ref{table2}, several other high-significance stellar groups were discovered that are not associated with known substructures (see Figures \ref{figure6} $-$ \ref{figure7}. With the exception of HAS, the other five stellar groups are characterized by disk-like dynamics and low-$\alpha$ abundances. Specifically, all five groups exhibit velocities with $V_{\phi}\,\sim\,200\,\mathrm{km\,s^{-1}}\, \mathrm{and}\, V_{r}\,,V_{z}\,\mathrm{centered\, around} \,0\,\mathrm{km\,s^{-1}}$, while HAS exhibits higher radial velocity with $|V_{r}|\,\sim\,150\,\mathrm{km\,s^{-1}}$ as shown in Figure \ref{figure6.5}. Additionally, during the second clustering process, the above five known substructures (Thamnos, GSE, Sgr, MWTD, LMS-1) were also retrieved, with new metal-rich member stars identified for GSE and Sgr. Consequently, GSE and Sgr are shown again in kinematic and chemical spaces (see Figures \ref{figure6} $-$ \ref{figure7}) alongside the groups/substructures in Table \ref{table2}.

\par $\bullet $ Aleph:it was first discovered by \citet{Naidu2020evidence} and characterized by its metal-rich ([Fe/H] $>-0.8$ dex) and relatively $\alpha$-poor ([$\alpha$/Fe] $<$ 0.27 dex) nature with significant vertical action ($J_{z} > J_{R}$). \citet{Naidu2020evidence} also showed that Aleph and the globular cluster Palomar 1 share many chemodynamical properties in common, suggesting a common origin. Among the six newly identified groups/substructures in Table \ref{table2}, Aleph is distinguishable according to \citet{Naidu2020evidence}: $175\,<V_{\phi}\,<\,300 (\mathrm{km\,s^{-1}})$, $|V_{r}|<\,75\mathrm{km\,s^{-1}}\,$, [Fe/H] $>\,-0.8 $ dex, [Mg/Fe] $< 0.27$ dex. Finally, we obtained a valid group with 71 stars. Of these, 50 stars exhibit the condition $J_{z} - J_{R} >\,0$, which is an acceptable number for considering this group as a representative of Aleph. 

\par $\bullet $ Gaia-Sausage-Enceladus: during the second clustering process, 1302 GSE stars were identified. Notably, the metal-rich GSE stars overlap with both high-$\alpha$ and low-$\alpha$ disk in the [Fe/H]-[Mg/Fe] plane. Note that the identified GSE could be contaminated by the Splash stars due to the kinematic similarities. Similar to the results in Table \ref{tab:idectification1}, some GSE stars were separated from the MWTD in a high-significance group and overlap with MWTD in chemical space. These observations imply a possibility that some ancient Galactic disk stars or gas clouds may have been perturbed by the gravitational influence of GSE, resulting in changes to their kinematic properties. Ancient disk stars, significantly affected by GSE, along with newly formed stars in the gas clouds, are expected to exhibit kinematic characteristics similar to those of GSE, and as a result, may be classified as GSE members in our study. It appears reasonable for the Splash identified as GSE members, because the orbits of the Splash stars have been much radialized \citep{Belokurov2020MNRAS}. The less influenced part of metal-poor ancient thick disk stars would exhibit MWTD-like kinematic features, which supports the hypothesis of \citet{Carollo2019evidence} that MWTD originates from ancient thick disk stars and was heated by the GSE-like merger. Therefore, the GSE merger likely had not yet finished even after the formation of the low-$\alpha$ disk. According to \citet{Tononi2019effects}, the thin disk (low-$\alpha$ disk) has been forming for 8–9 Gyr, while the thick disk (high-$\alpha$ disk) is at least 1.6 Gyr older \citep{Kilic2017the}. Hence, the GSE was likely accreted $\sim$8–9 Gyr ago, as supported by \citet{Belokurov2018Co}. However, the above hypothesis still needs to be verified through further studies.

\par $\bullet $ Sagittarius Stream: there are 48 high-significance stars belong to this substructure in the second clustering result. The above absent upside-down $\alpha$-``knee'' of Sgr now appears at [Fe/H] $\sim\,-0.8$ dex, where the [Mg/Fe] stops decreasing and shows an upside-down ``knee'' \citep{Hasselquist2021,Horta2023the}. It signifies [Mg/Fe] enrichment and indicates a burst in SF during the interaction with MW \citep{Hasselquist2021,Horta2023the}. Thus the occurrence of SNe II  would lead to a rapid rise in ISM enrichment of elements like Mg over a short timescale ($\sim 10^7$ years). Since Fe is mainly produced by SN Ia over a longer timescale ($\sim 10^8$-$10^9$ years), Fe enrichment lags, causing a sudden increase in [Mg/Fe] \citep{Hasselquist2021,Horta2023the}.

\par $\bullet $ HAS: this is a valid group of stars consisting of 102 member stars. HAS exhibits kinematic and chemical distributions within the following ranges: $E\,\sim\,[-5.0,\,-2.0]$$\times 10^{4}$ km$^2$ s$^-$$^2$ , $L_{z}\,\sim\,[500,\,2000]\,\mathrm{kpc}\,\mathrm{km\,s^{-1}}$, $J_{r}\,\sim\,[400,\,900]\,\mathrm{kpc}\,\mathrm{km\,s^{-1}}$, $J_{z}\,\sim\,[0,\,200]\,\mathrm{kpc}\,\mathrm{km\,s^{-1}}$, [Fe/H]$\,\sim\,$[$-$1.1, $-$0.1] dex, [Mg/Fe]$\,\sim\,$[0.2, 0.4] dex, [Al/Fe]$\,\sim\,$[0.1, 0.3] dex, $|V_{r}|\,\sim\,150\,\mathrm{km\,s^{-1}}$. Notably, HAS overlaps with the high-$\alpha$ disk (thick disk), suggesting the possibility that it may belong to the Galactic disk or the Splash \citep{Belokurov2020MNRAS}. However, the distribution of $V_{\phi}$ distinguishes HAS (with most stars having $V_{\phi} > 100\,\mathrm{km\,s^{-1}}$) from the Splash (with $V_{\phi} < 100\,\mathrm{km\,s^{-1}}$) \citep{Belokurov2020MNRAS}. Additionally, the high $|V_{r}|$ (approximately $150\,\mathrm{km\,s^{-1}}$), eccentricity ($e > 0.5$), and relatively low $\overline{V_{\phi}}$ (with a mean rotational velocity of $\sim\,156\,\mathrm{km\,s^{-1}}$) are indicative of halo-like orbits. Interestingly, we noticed that the member stars of HAS with positive \( V_{r} \) can be classified as members of the Nyx stream, as identified by \citet{Necib2020evidence}.  The remaining members of HAS conform to the definition of Nyx-2 (having opposite radial velocities compared to Nyx) and are proposed to share the same origin with Nyx \citep{Necib2020evidence}. As emphasized by \citet{Necib2020evidence}, the chemical similarities of the HAS member stars exhibited by our result would provide further support for the hypothesis that Nyx and Nyx-2 have a common origin. Thus, according to previous studies \citep[e.g.,][]{Bonaca2017ApJ, Belokurov2020MNRAS, Zucker2021ApJ}, the halo-like orbits of HAS/Nyx suggest that they may have originated from the thick disk and likely undergone significant radial migration and heating due to a merger event.

\par $\bullet $ LAS: it contains 80 member stars with a significance greater than $5\sigma$. LAS exhibits kinematic and chemical distribution in the following ranges: $E\,\sim\,[-5.0,\,-3.0]$$\times 10^{4}$ km$^2$ s$^-$$^2$ , $L_{z}\,\sim\,[1400,\,2200]\,\mathrm{kpc}\,\mathrm{km\,s^{-1}}$, $J_{r}\,\sim\,[0,\,100]\,\mathrm{kpc}\,\mathrm{km\,s^{-1}}$, $J_{z}\,\sim\,[0,\,15]\,\mathrm{kpc}\,\mathrm{km\,s^{-1}}$, [Fe/H]$\,\sim\,$[$-$1.0, 0] dex, [Mg/Fe]$\,\sim\,$[$-$0.6, $-$0.1] dex, [Al/Fe]$\,\sim\,$[$-$0.2, 0.2] dex. As exhibited in Figure \ref{figure7}, despite its disk-like dynamics, LAS has a notably lower $\alpha$ abundance and relatively rich metallicity (most of member stars have [Fe/H] $>\,-$0.5 dex), distinguishing it from known substructures. However, as demonstrated in the bottom-right panel of Figure \ref{figure7}, the LAS overlaps with the in-situ disk, thereby supporting our classification of this substructure as a distinct low-$\alpha$ substructure of the Galactic disk.

\par $\bullet $ NSTC-1: this group consists of 37 member stars, displaying disk-like kinematics and chemical distribution: $E\,\sim\,[-6.2,\,-2.5]$$\,\times 10^{4}$ km$^2$ s$^-$$^2$ , $L_{z}\,\sim\,[1.2,\,2.2]\,\times 10^{3}\,\mathrm{kpc}\,\mathrm{km\,s^{-1}}$, $J_{r}\,<\,100\,\mathrm{kpc}\,\mathrm{km\,s^{-1}}$, $J_{z}\,<\,20\,\mathrm{kpc}\,\mathrm{km\,s^{-1}}$, [Fe/H]$\,\sim\,$[$-$1.15, 0] dex, [Mg/Fe]$\,<\,0$ dex, [Al/Fe]$\,\sim\,$[$-$0.8, $-$0.05] dex. While the distribution of NSTC-1 in the [Fe/H]-[Mg/Fe] plane initially suggested an association with Aleph, but it is excluded by the difference in [Al/Fe] (shown in Figure \ref{figure7}). The [Al/Fe] difference also distinguishes NSTC-1 from the Galactic disk. Furthermore, its notably low [$\alpha$/Fe] and $V_{\phi}$ $>\,195 \,\mathrm{km\,s^{-1}}$ argue against its identification as Nyx. In [Fe/H]-[Al/Fe] plane, NSTC-1 is proximate to Sgr but diverges in the $(\mathbf{J},E)$ distribution. And according to \citet{Hasselquist2021}, NSTC-1 is too close ($d_{\odot}\,<4$ kpc) to be Sgr, confirming their separate identities. Therefore, we identify NSTC-1 as a newly discovered substructure candidate, and its striking resemblance to dwarf galaxy like Sgr in the [Fe/H]-[Mg/Fe] plane suggests that NSTC-1 may have originated ex-situ.

\par $\bullet $ NSTC-2: containing 32 member stars, this group exhibits a high significance exceeding 5$\sigma$. In addition to its disk-like velocities, it shows kinematic and chemical distributions within the following ranges: $E\,\sim\,[-6.0,\,-2.5]$$\,\times 10^{4}$ km$^2$ s$^-$$^2$ , $L_{z}\,\sim\,[1.2,\,2.2]\,\times 10^{3}\,\mathrm{kpc}\,\mathrm{km\,s^{-1}}$, $J_{r}\,<\,100\,\mathrm{kpc}\,\mathrm{km\,s^{-1}}$, $J_{z}\,<\,40\,\mathrm{kpc}\,\mathrm{km\,s^{-1}}$, [Fe/H]$\,\sim\,$[$-$1.0, 0] dex , [Mg/Fe]$\,\sim\,$[0, 0.2] dex, [Al/Fe]$\,\sim\,$[$-$0.7, $-$0.2] dex. Its metallicity predominantly ranges between [$-$1.0, $-$0.6], with most member stars exhibiting [Al/Fe] $<\,-$0.3. In $(L_{z},E)$ plane, NSTC-2 exhibits Aleph-like distribution with $e\,<$ 0.3 \citep{Naidu2020evidence}. However, NSTC-2 shows lower [Al/Fe] than Aleph as shown in Figure \ref{figure7}, which is close to the in-situ disk \citep{Horta2023the}, thereby excluding their association. Despite similar distribution in $(L_{z},E)$ plane compared to Nyx, NSTC-2 has higher $V_{\phi}$ ($> 195 \,\mathrm{km,s^{-1}}$) and lower [$\alpha$/Fe]. Based on these facts, we conclude that NSTC-2 is a distinct substructure, separating from both Aleph and Nyx.

\par $\bullet $ NSTC-3: it comprises 40 member stars and exhibits a significance exceeding $5\sigma$. Its kinematic and chemical distributions are as follows: $E\,\sim\,[-2.5,\,-0.9]$$\times 10^{4}$ km$^2$ s$^-$$^2$ , $L_{z}\,\sim\,[2.5,\,3.8]\,\times 10^{3}\,\mathrm{kpc}\,\mathrm{km\,s^{-1}}$, $J_{r}\,\sim\,[0,\,280]\,\mathrm{kpc}\,\mathrm{km\,s^{-1}}$, $J_{z}\,\sim\,[0,\,100]\,\mathrm{kpc}\,\mathrm{km\,s^{-1}}$, [Fe/H]$\,\sim\,$[$-$0.95, $-$0.35] dex, [Mg/Fe]$\,\sim\,$[$-$0.05, 0.2] dex, [Al/Fe]$\,\sim\,$[$-$0.6, $-$0.15] dex. Similar to NSTC-2, NSTC-3 is distinct from both Aleph and Nyx. Therefore, we consider that the NSTC-3 comprises a distinct group of stars, corresponding to a newly identified substructure.


\par 
Low [Al/Fe] and [Mg/Fe] has been postulated to be indicative of accretion events \citep[e.g.][]{Hasselquist2021}, as exemplified by the characteristics of the GSE and Sagittarius Stream in Figure \ref{figure4} and Figure \ref{figure5}. The result is consistent with recent studies, which have collectively attested to the low [Al/Fe] in the satellite galaxies of the MW in contrast to populations of in-situ origin \citep[e.g.,][]{Hawkins2015using, das2020, Horta2021evidence}. In our work, we adopt [Al/Fe] $<\,-$0.15 dex to identify low-Al stars as possibly originating ex-situ. Observed abundance trends in massive dwarf satellites, typically with [Al/Fe] $<\,-0.1$ dex, support this hypothesis \citep{Hasselquist2021,Belokurov2022}. Element Mn is an excellent pristine tracer of Type $\mathrm{Ia}$ SNe, as iron may originate from other formation channels, making the [Mg/Mn] ratio a reliable indicator for tracing $\alpha$-poor and $\alpha$-rich populations \citep{das2020}. Consequently, the [Al/Fe]-[Mg/Mn] plane effectively distinguishes high-$\alpha$, thick-disk stars from high-$\alpha$, accreted stars \citep{Hawkins2015using, das2020}. In \citet{das2020} and \citet{Horta2021evidence}, a region centered at ([Al/Fe], [Mg/Mn]) $\sim$ ($-$0.2, 0.5) is classified as part of the accreted halo. In our study, this region is defined by  [Mg/Mn] $>$ 0.1 dex and [Mg/Mn] $>$ 2.6$\times$[Al/Fe] + 0.45. Similar to the approach of \citet{Feuillet2022An}, the black-dashed line in the right panels of Figure \ref{figure7} is defined using the criteria outlined above, along with a threshold of [Al/Fe] = $-$0.15 dex. Therefore, the distributions of the three NSTC groups in Figure \ref{figure7} suggest that they are possibly accreted populations.

\par  Considering the fact that in some cases, an infalling dwarf galaxy can interact dynamically with the stellar disk and be preferentially dragged into the Galactic plane, leaving behind tidal debris that rotates with the disk [\citep{Quinn1986ApJ,Walker1996ApJ}, we propose that the NSTCs are likely to be substructures located on Galactic disk ($|z\,<\,2|$ kpc) and originate from dwarf galaxies accreted by the MW. \citet{Horta2021evidence} have shown that dwarf galaxies can also evolve to exhibit characteristics in the [Al/Fe]–[Mg/Mn] plane that are similar to those of NSTCs (low [Mg/Mn] and relatively low [Al/Fe]). Consequently, NSTCs may be associated with recent accretion events. Moreover, as discussed by \citet{Donlon2023ApJ}, material from a high-[Fe/H], low-[$\alpha$/Fe] dwarf galaxy that was recently accreted would possess high energy, which is consistent with the kinematic and chemical properties observed in NSTCs. This provides further evidence for the ex-situ origin of NSTCs and their potential connection to recent accretion events, but this remains to be confirmed through additional study.

\section{Summary and Conclusion}\label{5}
Based on the metal-poor stars ( [Fe/H] $<\,-$0.8 dex) in the APOGEE DR17, we adopted the StarGo algorithm to identify 42 groups and categorize them into five established substructures: GSE, LMS-1/Wukong, Thamnos, Sgr and MWTD. Once the constraint on [Fe/H] was removed, the metal-rich stars of the GSE were identified. These stars suggest that the accretion event had not been finished yet even after the formation of thin disk. Member stars of GSE separated from Thamnos and MWTD during the decline of $u_{thr}$ may also suggest that the GSE may have formed through multiple accretion events. 
\par Beyond these known substructures, two additional group of stars: UDG and NSTC were also identified, neither of which is linked to any previously recognized substructure. Among these two substructures, the UDG is proposed to be associated with Aleph. To further characterize the NSTC, we reapplied the clustering process without the prior [Fe/H] criteria. This approach confirmed the presence of Aleph and revealed five additional substructures, NSTC-1,2,3, HAS and LAS. The LAS is considered as a group of stars in the Galactic thin disk. While the HAS is proposed to be a combination of Nyx and Nyx-2, which also supports the hypothesis of a common origin for Nyx and Nyx-2 \citep{Necib2020evidence}. Additionally, low [Al/Fe] ($<-0.15$ dex) and relatively low [Mg/Fe] ($<0.2$ dex) indicates the similarities between satellite galaxies and NSTCs. Given the relatively low [$\alpha$/Fe] ($<0.2$ dex), relatively high [Fe/H] ($>-1.0$ dex) and $E$ ($>\,-6\times10^{4} \,\mathrm{km^{2}s^{-2}}$) of the three NSTC substructures, we consider that probably they have ex-situ origins with recent accretion events.
\par It is important to note that the above analysis is based on clustering results in kinematic and chemical spaces. Therefore, the progenitors of the substructures identified in these spaces remain uncertain \citep[e.g.,][]{Mori2024eas} and detailed properties of these progenitors will be investigated in the future work through high-resolution zoom in simulations with stellar abundance tracking \citep[e.g.,][]{Grand2017}. Furthermore, the newly identified substructures in this study may require additional validation through future astrometric surveys, particularly with the forthcoming Gaia Data Release 4 (DR4) and photometry from CSST and LSST, spectroscopy from WEAVE, 4MOST, etc.

\section{acknowledgments}
We thank especially the referee for insightful comments and suggestions, which have improved the paper significantly. This work was supported by National Key R\&D Program of China No. 2024YFA1611900, and the National Natural Science Foundation of China (NSFC Nos. 11973042, 11973052). In this work, we make use of data from $\href{https://www.sdss4.org/dr17/data_access/volume/}{Apache\,Point\,Observatory\,Galaxy\,Evolution\,Experiment\,Data\,Release\,17}$. Funding for the Sloan Digital Sky 
Survey IV has been provided by the 
Alfred P. Sloan Foundation, the U.S. 
Department of Energy Office of 
Science, and the Participating 
Institutions. 

SDSS-IV acknowledges support and 
resources from the Center for High 
Performance Computing  at the 
University of Utah. The SDSS 
website is www.sdss4.org.

SDSS-IV is managed by the 
Astrophysical Research Consortium 
for the Participating Institutions 
of the SDSS Collaboration including 
the Brazilian Participation Group, 
the Carnegie Institution for Science, 
Carnegie Mellon University, Center for 
Astrophysics | Harvard \& 
Smithsonian, the Chilean Participation 
Group, the French Participation Group, 
Instituto de Astrof\'isica de 
Canarias, The Johns Hopkins 
University, Kavli Institute for the 
Physics and Mathematics of the 
Universe (IPMU) / University of 
Tokyo, the Korean Participation Group, 
Lawrence Berkeley National Laboratory, 
Leibniz Institut f\"ur Astrophysik 
Potsdam (AIP),  Max-Planck-Institut 
f\"ur Astronomie (MPIA Heidelberg), 
Max-Planck-Institut f\"ur 
Astrophysik (MPA Garching), 
Max-Planck-Institut f\"ur 
Extraterrestrische Physik (MPE), 
National Astronomical Observatories of 
China, New Mexico State University, 
New York University, University of 
Notre Dame, Observat\'ario 
Nacional / MCTI, The Ohio State 
University, Pennsylvania State 
University, Shanghai 
Astronomical Observatory, United 
Kingdom Participation Group, 
Universidad Nacional Aut\'onoma 
de M\'exico, University of Arizona, 
University of Colorado Boulder, 
University of Oxford, University of 
Portsmouth, University of Utah, 
University of Virginia, University 
of Washington, University of 
Wisconsin, Vanderbilt University, 
and Yale University.

\bibliography{lzc}{}

\begin{thebibliography}{}
\expandafter\ifx\csname natexlab\endcsname\relax\def\natexlab#1{#1}\fi
\providecommand{\url}[1]{\href{#1}{#1}}
\providecommand{\dodoi}[1]{doi:~\href{http://doi.org/#1}{\nolinkurl{#1}}}
\providecommand{\doeprint}[1]{\href{http://ascl.net/#1}{\nolinkurl{http://ascl.net/#1}}}
\providecommand{\doarXiv}[1]{\href{https://arxiv.org/abs/#1}{\nolinkurl{https://arxiv.org/abs/#1}}}

\bibitem[{{Abdurro'uf} {et~al.}(2022){Abdurro'uf}, {Accetta}, {Aerts}, {Silva
  Aguirre}, {Ahumada}, {Ajgaonkar}, {Filiz Ak}, {Alam}, {Allende Prieto},
  {Almeida}, {Anders}, {Anderson}, {Andrews}, {Anguiano}, {Aquino-Ort{\'\i}z},
  {Arag{\'o}n-Salamanca}, {Argudo-Fern{\'a}ndez}, {Ata}, {Aubert},
  {Avila-Reese}, {Badenes}, {Barb{\'a}}, {Barger}, {Barrera-Ballesteros},
  {Beaton}, {Beers}, {Belfiore}, {Bender}, {Bernardi}, {Bershady}, {Beutler},
  {Bidin}, {Bird}, {Bizyaev}, {Blanc}, {Blanton}, {Boardman}, {Bolton},
  {Boquien}, {Borissova}, {Bovy}, {Brandt}, {Brown}, {Brownstein}, {Brusa},
  {Buchner}, {Bundy}, {Burchett}, {Bureau}, {Burgasser}, {Cabang}, {Campbell},
  {Cappellari}, {Carlberg}, {Wanderley}, {Carrera}, {Cash}, {Chen}, {Chen},
  {Cherinka}, {Chiappini}, {Choi}, {Chojnowski}, {Chung}, {Clerc}, {Cohen},
  {Comerford}, {Comparat}, {da Costa}, {Covey}, {Crane}, {Cruz-Gonzalez},
  {Culhane}, {Cunha}, {Dai}, {Damke}, {Darling}, {Davidson}, {Davies},
  {Dawson}, {De Lee}, {Diamond-Stanic}, {Cano-D{\'\i}az}, {S{\'a}nchez},
  {Donor}, {Duckworth}, {Dwelly}, {Eisenstein}, {Elsworth}, {Emsellem},
  {Eracleous}, {Escoffier}, {Fan}, {Farr}, {Feng}, {Fern{\'a}ndez-Trincado},
  {Feuillet}, {Filipp}, {Fillingham}, {Frinchaboy}, {Fromenteau}, {Galbany},
  {Garc{\'\i}a}, {Garc{\'\i}a-Hern{\'a}ndez}, {Ge}, {Geisler}, {Gelfand},
  {G{\'e}ron}, {Gibson}, {Goddy}, {Godoy-Rivera}, {Grabowski}, {Green},
  {Greener}, {Grier}, {Griffith}, {Guo}, {Guy}, {Hadjara}, {Harding},
  {Hasselquist}, {Hayes}, {Hearty}, {Hern{\'a}ndez}, {Hill}, {Hogg},
  {Holtzman}, {Horta}, {Hsieh}, {Hsu}, {Hsu}, {Huber}, {Huertas-Company},
  {Hutchinson}, {Hwang}, {Ibarra-Medel}, {Chitham}, {Ilha}, {Imig}, {Jaekle},
  {Jayasinghe}, {Ji}, {Johnson}, {Jones}, {J{\"o}nsson}, {Katkov}, {Khalatyan},
  {Kinemuchi}, {Kisku}, {Knapen}, {Kneib}, {Kollmeier}, {Kong}, {Kounkel},
  {Kreckel}, {Krishnarao}, {Lacerna}, {Lane}, {Langgin}, {Lavender}, {Law},
  {Lazarz}, {Leung}, {Leung}, {Lewis}, {Li}, {Li}, {Lian}, {Liang}, {Lin},
  {Lin}, {Lin}, {Lintott}, {Long}, {Longa-Pe{\~n}a}, {L{\'o}pez-Cob{\'a}},
  {Lu}, {Lundgren}, {Luo}, {Mackereth}, {de la Macorra}, {Mahadevan},
  {Majewski}, {Manchado}, {Mandeville}, {Maraston}, {Margalef-Bentabol},
  {Masseron}, {Masters}, {Mathur}, {McDermid}, {Mckay}, {Merloni},
  {Merrifield}, {Meszaros}, {Miglio}, {Di Mille}, {Minniti}, {Minsley},
  {Monachesi}, {Moon}, {Mosser}, {Mulchaey}, {Muna}, {Mu{\~n}oz}, {Myers},
  {Myers}, {Nadathur}, {Nair}, {Nandra}, {Neumann}, {Newman}, {Nidever},
  {Nikakhtar}, {Nitschelm}, {O'Connell}, {Garma-Oehmichen}, {Luan Souza de
  Oliveira}, {Olney}, {Oravetz}, {Ortigoza-Urdaneta}, {Osorio}, {Otter},
  {Pace}, {Padilla}, {Pan}, {Pan}, {Parikh}, {Parker}, {Peirani}, {Pe{\~n}a
  Ram{\'\i}rez}, {Penny}, {Percival}, {Perez-Fournon}, {Pinsonneault},
  {Poidevin}, {Poovelil}, {Price-Whelan}, {B{\'a}rbara de Andrade Queiroz},
  {Raddick}, {Ray}, {Rembold}, {Riddle}, {Riffel}, {Riffel}, {Rix}, {Robin},
  {Rodr{\'\i}guez-Puebla}, {Roman-Lopes}, {Rom{\'a}n-Z{\'u}{\~n}iga}, {Rose},
  {Ross}, {Rossi}, {Rubin}, {Salvato}, {S{\'a}nchez}, {S{\'a}nchez-Gallego},
  {Sanderson}, {Santana Rojas}, {Sarceno}, {Sarmiento}, {Sayres}, {Sazonova},
  {Schaefer}, {Schiavon}, {Schlegel}, {Schneider}, {Schultheis}, {Schwope},
  {Serenelli}, {Serna}, {Shao}, {Shapiro}, {Sharma}, {Shen}, {Shetrone}, {Shu},
  {Simon}, {Skrutskie}, {Smethurst}, {Smith}, {Sobeck}, {Spoo}, {Sprague},
  {Stark}, {Stassun}, {Steinmetz}, {Stello}, {Stone-Martinez},
  {Storchi-Bergmann}, {Stringfellow}, {Stutz}, {Su}, {Taghizadeh-Popp},
  {Talbot}, {Tayar}, {Telles}, {Teske}, {Thakar}, {Theissen}, {Tkachenko},
  {Thomas}, {Tojeiro}, {Hernandez Toledo}, {Troup}, {Trump}, {Trussler},
  {Turner}, {Tuttle}, {Unda-Sanzana}, {V{\'a}zquez-Mata}, {Valentini},
  {Valenzuela}, {Vargas-Gonz{\'a}lez}, {Vargas-Maga{\~n}a}, {Alfaro},
  {Villanova}, {Vincenzo}, {Wake}, {Warfield}, {Washington}, {Weaver},
  {Weijmans}, {Weinberg}, {Weiss}, {Westfall}, {Wild}, {Wilde}, {Wilson},
  {Wilson}, {Wilson}, {Wolf}, {Wood-Vasey}, {Yan}, {Zamora}, {Zasowski},
  {Zhang}, {Zhao}, {Zheng}, {Zheng}, \& {Zhu}}]{Abdurro2022ApJthe}
{Abdurro'uf}, {Accetta}, K., {Aerts}, C., {et~al.} 2022, \apjs, 259, 35,
  \dodoi{10.3847/1538-4365/ac4414}

\bibitem[{{Amarante} {et~al.}(2022){Amarante}, {Debattista}, {Beraldo e Silva},
  {Laporte}, \& {Deg}}]{Amarante2022ApJ}
{Amarante}, J. A.~S., {Debattista}, V.~P., {Beraldo e Silva}, L., {Laporte}, C.
  F.~P., \& {Deg}, N. 2022, \apj, 937, 12, \dodoi{10.3847/1538-4357/ac8b0d}

\bibitem[{{Antoja} {et~al.}(2020){Antoja}, {Ramos}, {Mateu}, {Helmi}, {Anders},
  {Jordi}, \& {Carballo-Bello}}]{Antoja2020an}
{Antoja}, T., {Ramos}, P., {Mateu}, C., {et~al.} 2020, \aap, 635, L3,
  \dodoi{10.1051/0004-6361/201937145}

\bibitem[{{Barb{\'a}} {et~al.}(2019){Barb{\'a}}, {Minniti}, {Geisler},
  {Alonso-Garc{\'\i}a}, {Hempel}, {Monachesi}, {Arias}, \&
  {G{\'o}mez}}]{Barb2019as}
{Barb{\'a}}, R.~H., {Minniti}, D., {Geisler}, D., {et~al.} 2019, \apjl, 870,
  L24, \dodoi{10.3847/2041-8213/aaf811}

\bibitem[{{Belokurov} {et~al.}(2018){Belokurov}, {Erkal}, {Evans}, {Koposov},
  \& {Deason}}]{Belokurov2018Co}
{Belokurov}, V., {Erkal}, D., {Evans}, N.~W., {Koposov}, S.~E., \& {Deason},
  A.~J. 2018, \mnras, 478, 611, \dodoi{10.1093/mnras/sty982}

\bibitem[{Belokurov \& Kravtsov(2022)}]{Belokurov2022}
Belokurov, V., \& Kravtsov, A. 2022, \mnras, 514, 689,
  \dodoi{10.1093/mnras/stac1267}

\bibitem[{{Belokurov} {et~al.}(2020){Belokurov}, {Sanders}, {Fattahi}, {Smith},
  {Deason}, {Evans}, \& {Grand}}]{Belokurov2020MNRAS}
{Belokurov}, V., {Sanders}, J.~L., {Fattahi}, A., {et~al.} 2020, \mnras, 494,
  3880, \dodoi{10.1093/mnras/staa876}

\bibitem[{{Belokurov} {et~al.}(2006){Belokurov}, {Zucker}, {Evans}, {Gilmore},
  {Vidrih}, {Bramich}, {Newberg}, {Wyse}, {Irwin}, {Fellhauer}, {Hewett},
  {Walton}, {Wilkinson}, {Cole}, {Yanny}, {Rockosi}, {Beers}, {Bell},
  {Brinkmann}, {Ivezi{\'c}}, \& {Lupton}}]{Belokurov2006the}
{Belokurov}, V., {Zucker}, D.~B., {Evans}, N.~W., {et~al.} 2006, \apjl, 642,
  L137, \dodoi{10.1086/504797}

\bibitem[{Bennett \& Bovy(2018)}]{Bennett2018}
Bennett, M., \& Bovy, J. 2018, \mnras, 482, 1417, \dodoi{10.1093/mnras/sty2813}

\bibitem[{{Bonaca} {et~al.}(2017){Bonaca}, {Conroy}, {Wetzel}, {Hopkins}, \&
  {Kere{\v{s}}}}]{Bonaca2017ApJ}
{Bonaca}, A., {Conroy}, C., {Wetzel}, A., {Hopkins}, P.~F., \& {Kere{\v{s}}},
  D. 2017, \apj, 845, 101, \dodoi{10.3847/1538-4357/aa7d0c}

\bibitem[{{Bovy}(2015)}]{Bovy2015}
{Bovy}, J. 2015, \apjs, 216, 29, \dodoi{10.1088/0067-0049/216/2/29}

\bibitem[{{Carlin} {et~al.}(2018){Carlin}, {Sheffield}, {Cunha}, \&
  {Smith}}]{Carlin2018chemical}
{Carlin}, J.~L., {Sheffield}, A.~A., {Cunha}, K., \& {Smith}, V.~V. 2018,
  \apjl, 859, L10, \dodoi{10.3847/2041-8213/aac3d8}

\bibitem[{{Carollo} {et~al.}(2019){Carollo}, {Chiba}, {Ishigaki}, {Freeman},
  {Beers}, {Lee}, {Tissera}, {Battistini}, \& {Primas}}]{Carollo2019evidence}
{Carollo}, D., {Chiba}, M., {Ishigaki}, M., {et~al.} 2019, \apj, 887, 22,
  \dodoi{10.3847/1538-4357/ab517c}

\bibitem[{Chiba \& Beers(2000)}]{Chiba_2000}
Chiba, M., \& Beers, T.~C. 2000, \aj, 119, 2843, \dodoi{10.1086/301409}

\bibitem[{{Cui} {et~al.}(2012){Cui}, {Zhao}, {Chu}, {Li}, {Li}, {Zhang}, {Su},
  {Yao}, {Wang}, {Xing}, {Li}, {Zhu}, {Wang}, {Gu}, {Luo}, {Xu}, {Zhang},
  {Liu}, {Zhang}, {Yang}, {Cao}, {Chen}, {Chen}, {Chen}, {Chen}, {Chu}, {Feng},
  {Gong}, {Hou}, {Hu}, {Hu}, {Hu}, {Jia}, {Jiang}, {Jiang}, {Jiang}, {Jin},
  {Li}, {Li}, {Li}, {Liu}, {Liu}, {Lu}, {Mao}, {Men}, {Qi}, {Qi}, {Shi},
  {Tang}, {Tao}, {Wang}, {Wang}, {Wang}, {Wang}, {Wang}, {Wang}, {Wang},
  {Wang}, {Wang}, {Wang}, {Wang}, {Wang}, {Xu}, {Xu}, {Yang}, {Yu}, {Yuan},
  {Yuan}, {Zhai}, {Zhang}, {Zhang}, {Zhang}, {Zhao}, {Zhou}, {Zhou}, {Zhu}, \&
  {Zou}}]{Cui2012RAA}
{Cui}, X.-Q., {Zhao}, Y.-H., {Chu}, Y.-Q., {et~al.} 2012, RAA, 12, 1197,
  \dodoi{10.1088/1674-4527/12/9/003}

\bibitem[{{Das} {et~al.}(2020){Das}, {Hawkins}, \& {Jofr{\'e}}}]{das2020}
{Das}, P., {Hawkins}, K., \& {Jofr{\'e}}, P. 2020, \mnras, 493, 5195,
  \dodoi{10.1093/mnras/stz3537}

\bibitem[{{De Silva} {et~al.}(2015){De Silva}, {Freeman}, {Bland-Hawthorn},
  {Martell}, {de Boer}, {Asplund}, {Keller}, {Sharma}, {Zucker}, {Zwitter},
  {Anguiano}, {Bacigalupo}, {Bayliss}, {Beavis}, {Bergemann}, {Campbell},
  {Cannon}, {Carollo}, {Casagrande}, {Casey}, {Da Costa}, {D'Orazi}, {Dotter},
  {Duong}, {Heger}, {Ireland}, {Kafle}, {Kos}, {Lattanzio}, {Lewis}, {Lin},
  {Lind}, {Munari}, {Nataf}, {O'Toole}, {Parker}, {Reid}, {Schlesinger},
  {Sheinis}, {Simpson}, {Stello}, {Ting}, {Traven}, {Watson}, {Wittenmyer},
  {Yong}, \& {{\v{Z}}erjal}}]{DeSilva2015GALAH}
{De Silva}, G.~M., {Freeman}, K.~C., {Bland-Hawthorn}, J., {et~al.} 2015,
  \mnras, 449, 2604, \dodoi{10.1093/mnras/stv327}

\bibitem[{{Di Matteo} {et~al.}(2019){Di Matteo}, {Haywood}, {Lehnert}, {Katz},
  {Khoperskov}, {Snaith}, {G{\'o}mez}, \& {Robichon}}]{DiMatteo2019A&A}
{Di Matteo}, P., {Haywood}, M., {Lehnert}, M.~D., {et~al.} 2019, \aap, 632, A4,
  \dodoi{10.1051/0004-6361/201834929}

\bibitem[{Donlon \& Newberg(2022)}]{Donlon2022ASO}
Donlon, T., \& Newberg, H.~J. 2022, \apj, 944.
\newblock \url{https://api.semanticscholar.org/CorpusID:253801946}

\bibitem[{{Donlon} \& {Newberg}(2023)}]{Donlon2023ApJ}
{Donlon}, T., \& {Newberg}, H.~J. 2023, \apj, 944, 169,
  \dodoi{10.3847/1538-4357/acb150}

\bibitem[{{Donor} {et~al.}(2020){Donor}, {Frinchaboy}, {Cunha}, {O'Connell},
  {Allende Prieto}, {Almeida}, {Anders}, {Beaton}, {Bizyaev}, {Brownstein},
  {Carrera}, {Chiappini}, {Cohen}, {Garc{\'\i}a-Hern{\'a}ndez}, {Geisler},
  {Hasselquist}, {J{\"o}nsson}, {Lane}, {Majewski}, {Minniti}, {Bidin}, {Pan},
  {Roman-Lopes}, {Sobeck}, \& {Zasowski}}]{Donor2020}
{Donor}, J., {Frinchaboy}, P.~M., {Cunha}, K., {et~al.} 2020, \aj, 159, 199,
  \dodoi{10.3847/1538-3881/ab77bc}

\bibitem[{{Feuillet} {et~al.}(2022){Feuillet}, {Feltzing}, {Sahlholdt}, \&
  {Bensby}}]{Feuillet2022An}
{Feuillet}, D.~K., {Feltzing}, S., {Sahlholdt}, C., \& {Bensby}, T. 2022, \apj,
  934, 21, \dodoi{10.3847/1538-4357/ac76ba}

\bibitem[{{Gaia Collaboration} {et~al.}(2016){Gaia Collaboration}, {Prusti},
  {de Bruijne}, {Brown}, {Vallenari}, {Babusiaux}, {Bailer-Jones}, {Bastian},
  {Biermann}, {Evans}, {Eyer}, {Jansen}, {Jordi}, {Klioner}, {Lammers},
  {Lindegren}, {Luri}, {Mignard}, {Milligan}, {Panem}, {Poinsignon},
  {Pourbaix}, {Randich}, {Sarri}, {Sartoretti}, {Siddiqui}, {Soubiran},
  {Valette}, {van Leeuwen}, {Walton}, {Aerts}, {Arenou}, {Cropper}, {Drimmel},
  {H{\o}g}, {Katz}, {Lattanzi}, {O'Mullane}, {Grebel}, {Holland}, {Huc},
  {Passot}, {Bramante}, {Cacciari}, {Casta{\~n}eda}, {Chaoul}, {Cheek}, {De
  Angeli}, {Fabricius}, {Guerra}, {Hern{\'a}ndez}, {Jean-Antoine-Piccolo},
  {Masana}, {Messineo}, {Mowlavi}, {Nienartowicz}, {Ord{\'o}{\~n}ez-Blanco},
  {Panuzzo}, {Portell}, {Richards}, {Riello}, {Seabroke}, {Tanga},
  {Th{\'e}venin}, {Torra}, {Els}, {Gracia-Abril}, {Comoretto},
  {Garcia-Reinaldos}, {Lock}, {Mercier}, {Altmann}, {Andrae}, {Astraatmadja},
  {Bellas-Velidis}, {Benson}, {Berthier}, {Blomme}, {Busso}, {Carry},
  {Cellino}, {Clementini}, {Cowell}, {Creevey}, {Cuypers}, {Davidson}, {De
  Ridder}, {de Torres}, {Delchambre}, {Dell'Oro}, {Ducourant}, {Fr{\'e}mat},
  {Garc{\'\i}a-Torres}, {Gosset}, {Halbwachs}, {Hambly}, {Harrison}, {Hauser},
  {Hestroffer}, {Hodgkin}, {Huckle}, {Hutton}, {Jasniewicz}, {Jordan},
  {Kontizas}, {Korn}, {Lanzafame}, {Manteiga}, {Moitinho}, {Muinonen},
  {Osinde}, {Pancino}, {Pauwels}, {Petit}, {Recio-Blanco}, {Robin}, {Sarro},
  {Siopis}, {Smith}, {Smith}, {Sozzetti}, {Thuillot}, {van Reeven}, {Viala},
  {Abbas}, {Abreu Aramburu}, {Accart}, {Aguado}, {Allan}, {Allasia},
  {Altavilla}, {{\'A}lvarez}, {Alves}, {Anderson}, {Andrei}, {Anglada Varela},
  {Antiche}, {Antoja}, {Ant{\'o}n}, {Arcay}, {Atzei}, {Ayache}, {Bach},
  {Baker}, {Balaguer-N{\'u}{\~n}ez}, {Barache}, {Barata}, {Barbier}, {Barblan},
  {Baroni}, {Barrado y Navascu{\'e}s}, {Barros}, {Barstow}, {Becciani},
  {Bellazzini}, {Bellei}, {Bello Garc{\'\i}a}, {Belokurov}, {Bendjoya},
  {Berihuete}, {Bianchi}, {Bienaym{\'e}}, {Billebaud}, {Blagorodnova},
  {Blanco-Cuaresma}, {Boch}, {Bombrun}, {Borrachero}, {Bouquillon}, {Bourda},
  {Bouy}, {Bragaglia}, {Breddels}, {Brouillet}, {Br{\"u}semeister},
  {Bucciarelli}, {Budnik}, {Burgess}, {Burgon}, {Burlacu}, {Busonero}, {Buzzi},
  {Caffau}, {Cambras}, {Campbell}, {Cancelliere}, {Cantat-Gaudin}, {Carlucci},
  {Carrasco}, {Castellani}, {Charlot}, {Charnas}, {Charvet}, {Chassat},
  {Chiavassa}, {Clotet}, {Cocozza}, {Collins}, {Collins}, {Costigan}, {Crifo},
  {Cross}, {Crosta}, {Crowley}, {Dafonte}, {Damerdji}, {Dapergolas}, {David},
  {David}, {De Cat}, {de Felice}, {de Laverny}, {De Luise}, {De March}, {de
  Martino}, {de Souza}, {Debosscher}, {del Pozo}, {Delbo}, {Delgado},
  {Delgado}, {di Marco}, {Di Matteo}, {Diakite}, {Distefano}, {Dolding}, {Dos
  Anjos}, {Drazinos}, {Dur{\'a}n}, {Dzigan}, {Ecale}, {Edvardsson}, {Enke},
  {Erdmann}, {Escolar}, {Espina}, {Evans}, {Eynard Bontemps}, {Fabre},
  {Fabrizio}, {Faigler}, {Falc{\~a}o}, {Farr{\`a}s Casas}, {Faye}, {Federici},
  {Fedorets}, {Fern{\'a}ndez-Hern{\'a}ndez}, {Fernique}, {Fienga}, {Figueras},
  {Filippi}, {Findeisen}, {Fonti}, {Fouesneau}, {Fraile}, {Fraser}, {Fuchs},
  {Furnell}, {Gai}, {Galleti}, {Galluccio}, {Garabato}, {Garc{\'\i}a-Sedano},
  {Gar{\'e}}, {Garofalo}, {Garralda}, {Gavras}, {Gerssen}, {Geyer}, {Gilmore},
  {Girona}, {Giuffrida}, {Gomes}, {Gonz{\'a}lez-Marcos},
  {Gonz{\'a}lez-N{\'u}{\~n}ez}, {Gonz{\'a}lez-Vidal}, {Granvik}, {Guerrier},
  {Guillout}, {Guiraud}, {G{\'u}rpide}, {Guti{\'e}rrez-S{\'a}nchez}, {Guy},
  {Haigron}, {Hatzidimitriou}, {Haywood}, {Heiter}, {Helmi}, {Hobbs},
  {Hofmann}, {Holl}, {Holland}, {Hunt}, {Hypki}, {Icardi}, {Irwin}, {Jevardat
  de Fombelle}, {Jofr{\'e}}, {Jonker}, {Jorissen}, {Julbe}, {Karampelas},
  {Kochoska}, {Kohley}, {Kolenberg}, {Kontizas}, {Koposov}, {Kordopatis},
  {Koubsky}, {Kowalczyk}, {Krone-Martins}, {Kudryashova}, {Kull}, {Bachchan},
  {Lacoste-Seris}, {Lanza}, {Lavigne}, {Le Poncin-Lafitte}, {Lebreton},
  {Lebzelter}, {Leccia}, {Leclerc}, {Lecoeur-Taibi}, {Lemaitre}, {Lenhardt},
  {Leroux}, {Liao}, {Licata}, {Lindstr{\o}m}, {Lister}, {Livanou}, {Lobel},
  {L{\"o}ffler}, {L{\'o}pez}, {Lopez-Lozano}, {Lorenz}, {Loureiro},
  {MacDonald}, {Magalh{\~a}es Fernandes}, {Managau}, {Mann}, {Mantelet},
  {Marchal}, {Marchant}, {Marconi}, {Marie}, {Marinoni}, {Marrese},
  {Marschalk{\'o}}, {Marshall}, {Mart{\'\i}n-Fleitas}, {Martino}, {Mary},
  {Matijevi{\v{c}}}, {Mazeh}, {McMillan}, {Messina}, {Mestre}, {Michalik},
  {Millar}, {Miranda}, {Molina}, {Molinaro}, {Molinaro}, {Moln{\'a}r},
  {Moniez}, {Montegriffo}, {Monteiro}, {Mor}, {Mora}, {Morbidelli}, {Morel},
  {Morgenthaler}, {Morley}, {Morris}, {Mulone}, {Muraveva}, {Musella},
  {Narbonne}, {Nelemans}, {Nicastro}, {Noval}, {Ord{\'e}novic},
  {Ordieres-Mer{\'e}}, {Osborne}, {Pagani}, {Pagano}, {Pailler}, {Palacin},
  {Palaversa}, {Parsons}, {Paulsen}, {Pecoraro}, {Pedrosa}, {Pentik{\"a}inen},
  {Pereira}, {Pichon}, {Piersimoni}, {Pineau}, {Plachy}, {Plum}, {Poujoulet},
  {Pr{\v{s}}a}, {Pulone}, {Ragaini}, {Rago}, {Rambaux}, {Ramos-Lerate},
  {Ranalli}, {Rauw}, {Read}, {Regibo}, {Renk}, {Reyl{\'e}}, {Ribeiro},
  {Rimoldini}, {Ripepi}, {Riva}, {Rixon}, {Roelens}, {Romero-G{\'o}mez},
  {Rowell}, {Royer}, {Rudolph}, {Ruiz-Dern}, {Sadowski}, {Sagrist{\`a}
  Sell{\'e}s}, {Sahlmann}, {Salgado}, {Salguero}, {Sarasso}, {Savietto},
  {Schnorhk}, {Schultheis}, {Sciacca}, {Segol}, {Segovia}, {Segransan},
  {Serpell}, {Shih}, {Smareglia}, {Smart}, {Smith}, {Solano}, {Solitro},
  {Sordo}, {Soria Nieto}, {Souchay}, {Spagna}, {Spoto}, {Stampa}, {Steele},
  {Steidelm{\"u}ller}, {Stephenson}, {Stoev}, {Suess}, {S{\"u}veges}, {Surdej},
  {Szabados}, {Szegedi-Elek}, {Tapiador}, {Taris}, {Tauran}, {Taylor},
  {Teixeira}, {Terrett}, {Tingley}, {Trager}, {Turon}, {Ulla}, {Utrilla},
  {Valentini}, {van Elteren}, {Van Hemelryck}, {van Leeuwen}, {Varadi},
  {Vecchiato}, {Veljanoski}, {Via}, {Vicente}, {Vogt}, {Voss}, {Votruba},
  {Voutsinas}, {Walmsley}, {Weiler}, {Weingrill}, {Werner}, {Wevers},
  {Whitehead}, {Wyrzykowski}, {Yoldas}, {{\v{Z}}erjal}, {Zucker}, {Zurbach},
  {Zwitter}, {Alecu}, {Allen}, {Allende Prieto}, {Amorim},
  {Anglada-Escud{\'e}}, {Arsenijevic}, {Azaz}, {Balm}, {Beck}, {Bernstein},
  {Bigot}, {Bijaoui}, {Blasco}, {Bonfigli}, {Bono}, {Boudreault}, {Bressan},
  {Brown}, {Brunet}, {Bunclark}, {Buonanno}, {Butkevich}, {Carret}, {Carrion},
  {Chemin}, {Ch{\'e}reau}, {Corcione}, {Darmigny}, {de Boer}, {de Teodoro}, {de
  Zeeuw}, {Delle Luche}, {Domingues}, {Dubath}, {Fodor}, {Fr{\'e}zouls},
  {Fries}, {Fustes}, {Fyfe}, {Gallardo}, {Gallegos}, {Gardiol}, {Gebran},
  {Gomboc}, {G{\'o}mez}, {Grux}, {Gueguen}, {Heyrovsky}, {Hoar}, {Iannicola},
  {Isasi Parache}, {Janotto}, {Joliet}, {Jonckheere}, {Keil}, {Kim},
  {Klagyivik}, {Klar}, {Knude}, {Kochukhov}, {Kolka}, {Kos}, {Kutka}, {Lainey},
  {LeBouquin}, {Liu}, {Loreggia}, {Makarov}, {Marseille}, {Martayan},
  {Martinez-Rubi}, {Massart}, {Meynadier}, {Mignot}, {Munari}, {Nguyen},
  {Nordlander}, {Ocvirk}, {O'Flaherty}, {Olias Sanz}, {Ortiz}, {Osorio},
  {Oszkiewicz}, {Ouzounis}, {Palmer}, {Park}, {Pasquato}, {Peltzer}, {Peralta},
  {P{\'e}turaud}, {Pieniluoma}, {Pigozzi}, {Poels}, {Prat}, {Prod'homme},
  {Raison}, {Rebordao}, {Risquez}, {Rocca-Volmerange}, {Rosen}, {Ruiz-Fuertes},
  {Russo}, {Sembay}, {Serraller Vizcaino}, {Short}, {Siebert}, {Silva},
  {Sinachopoulos}, {Slezak}, {Soffel}, {Sosnowska}, {Strai{\v{z}}ys}, {ter
  Linden}, {Terrell}, {Theil}, {Tiede}, {Troisi}, {Tsalmantza}, {Tur},
  {Vaccari}, {Vachier}, {Valles}, {Van Hamme}, {Veltz}, {Virtanen}, {Wallut},
  {Wichmann}, {Wilkinson}, {Ziaeepour}, \& {Zschocke}}]{Gaia2016A&A}
{Gaia Collaboration}, {Prusti}, T., {de Bruijne}, J.~H.~J., {et~al.} 2016,
  \aap, 595, A1, \dodoi{10.1051/0004-6361/201629272}

\bibitem[{{Gaia Collaboration} {et~al.}(2023){Gaia Collaboration}, {Vallenari},
  {Brown}, {Prusti}, {de Bruijne}, {Arenou}, {Babusiaux}, {Biermann},
  {Creevey}, {Ducourant}, {Evans}, {Eyer}, {Guerra}, {Hutton}, {Jordi},
  {Klioner}, {Lammers}, {Lindegren}, {Luri}, {Mignard}, {Panem}, {Pourbaix},
  {Randich}, {Sartoretti}, {Soubiran}, {Tanga}, {Walton}, {Bailer-Jones},
  {Bastian}, {Drimmel}, {Jansen}, {Katz}, {Lattanzi}, {van Leeuwen}, {Bakker},
  {Cacciari}, {Casta{\~n}eda}, {De Angeli}, {Fabricius}, {Fouesneau},
  {Fr{\'e}mat}, {Galluccio}, {Guerrier}, {Heiter}, {Masana}, {Messineo},
  {Mowlavi}, {Nicolas}, {Nienartowicz}, {Pailler}, {Panuzzo}, {Riclet}, {Roux},
  {Seabroke}, {Sordo}, {Th{\'e}venin}, {Gracia-Abril}, {Portell}, {Teyssier},
  {Altmann}, {Andrae}, {Audard}, {Bellas-Velidis}, {Benson}, {Berthier},
  {Blomme}, {Burgess}, {Busonero}, {Busso}, {C{\'a}novas}, {Carry}, {Cellino},
  {Cheek}, {Clementini}, {Damerdji}, {Davidson}, {de Teodoro}, {Nu{\~n}ez
  Campos}, {Delchambre}, {Dell'Oro}, {Esquej}, {Fern{\'a}ndez-Hern{\'a}ndez},
  {Fraile}, {Garabato}, {Garc{\'\i}a-Lario}, {Gosset}, {Haigron}, {Halbwachs},
  {Hambly}, {Harrison}, {Hern{\'a}ndez}, {Hestroffer}, {Hodgkin}, {Holl},
  {Jan{\ss}en}, {Jevardat de Fombelle}, {Jordan}, {Krone-Martins}, {Lanzafame},
  {L{\"o}ffler}, {Marchal}, {Marrese}, {Moitinho}, {Muinonen}, {Osborne},
  {Pancino}, {Pauwels}, {Recio-Blanco}, {Reyl{\'e}}, {Riello}, {Rimoldini},
  {Roegiers}, {Rybizki}, {Sarro}, {Siopis}, {Smith}, {Sozzetti}, {Utrilla},
  {van Leeuwen}, {Abbas}, {{\'A}brah{\'a}m}, {Abreu Aramburu}, {Aerts},
  {Aguado}, {Ajaj}, {Aldea-Montero}, {Altavilla}, {{\'A}lvarez}, {Alves},
  {Anders}, {Anderson}, {Anglada Varela}, {Antoja}, {Baines}, {Baker},
  {Balaguer-N{\'u}{\~n}ez}, {Balbinot}, {Balog}, {Barache}, {Barbato},
  {Barros}, {Barstow}, {Bartolom{\'e}}, {Bassilana}, {Bauchet}, {Becciani},
  {Bellazzini}, {Berihuete}, {Bernet}, {Bertone}, {Bianchi}, {Binnenfeld},
  {Blanco-Cuaresma}, {Blazere}, {Boch}, {Bombrun}, {Bossini}, {Bouquillon},
  {Bragaglia}, {Bramante}, {Breedt}, {Bressan}, {Brouillet}, {Brugaletta},
  {Bucciarelli}, {Burlacu}, {Butkevich}, {Buzzi}, {Caffau}, {Cancelliere},
  {Cantat-Gaudin}, {Carballo}, {Carlucci}, {Carnerero}, {Carrasco},
  {Casamiquela}, {Castellani}, {Castro-Ginard}, {Chaoul}, {Charlot}, {Chemin},
  {Chiaramida}, {Chiavassa}, {Chornay}, {Comoretto}, {Contursi}, {Cooper},
  {Cornez}, {Cowell}, {Crifo}, {Cropper}, {Crosta}, {Crowley}, {Dafonte},
  {Dapergolas}, {David}, {David}, {de Laverny}, {De Luise}, {De March}, {De
  Ridder}, {de Souza}, {de Torres}, {del Peloso}, {del Pozo}, {Delbo},
  {Delgado}, {Delisle}, {Demouchy}, {Dharmawardena}, {Di Matteo}, {Diakite},
  {Diener}, {Distefano}, {Dolding}, {Edvardsson}, {Enke}, {Fabre}, {Fabrizio},
  {Faigler}, {Fedorets}, {Fernique}, {Fienga}, {Figueras}, {Fournier},
  {Fouron}, {Fragkoudi}, {Gai}, {Garcia-Gutierrez}, {Garcia-Reinaldos},
  {Garc{\'\i}a-Torres}, {Garofalo}, {Gavel}, {Gavras}, {Gerlach}, {Geyer},
  {Giacobbe}, {Gilmore}, {Girona}, {Giuffrida}, {Gomel}, {Gomez},
  {Gonz{\'a}lez-N{\'u}{\~n}ez}, {Gonz{\'a}lez-Santamar{\'\i}a},
  {Gonz{\'a}lez-Vidal}, {Granvik}, {Guillout}, {Guiraud},
  {Guti{\'e}rrez-S{\'a}nchez}, {Guy}, {Hatzidimitriou}, {Hauser}, {Haywood},
  {Helmer}, {Helmi}, {Sarmiento}, {Hidalgo}, {Hilger}, {H{\l}adczuk}, {Hobbs},
  {Holland}, {Huckle}, {Jardine}, {Jasniewicz}, {Jean-Antoine Piccolo},
  {Jim{\'e}nez-Arranz}, {Jorissen}, {Juaristi Campillo}, {Julbe}, {Karbevska},
  {Kervella}, {Khanna}, {Kontizas}, {Kordopatis}, {Korn}, {K{\'o}sp{\'a}l},
  {Kostrzewa-Rutkowska}, {Kruszy{\'n}ska}, {Kun}, {Laizeau}, {Lambert},
  {Lanza}, {Lasne}, {Le Campion}, {Lebreton}, {Lebzelter}, {Leccia}, {Leclerc},
  {Lecoeur-Taibi}, {Liao}, {Licata}, {Lindstr{\o}m}, {Lister}, {Livanou},
  {Lobel}, {Lorca}, {Loup}, {Madrero Pardo}, {Magdaleno Romeo}, {Managau},
  {Mann}, {Manteiga}, {Marchant}, {Marconi}, {Marcos}, {Marcos Santos},
  {Mar{\'\i}n Pina}, {Marinoni}, {Marocco}, {Marshall}, {Martin Polo},
  {Mart{\'\i}n-Fleitas}, {Marton}, {Mary}, {Masip}, {Massari},
  {Mastrobuono-Battisti}, {Mazeh}, {McMillan}, {Messina}, {Michalik}, {Millar},
  {Mints}, {Molina}, {Molinaro}, {Moln{\'a}r}, {Monari}, {Mongui{\'o}},
  {Montegriffo}, {Montero}, {Mor}, {Mora}, {Morbidelli}, {Morel}, {Morris},
  {Muraveva}, {Murphy}, {Musella}, {Nagy}, {Noval}, {Oca{\~n}a}, {Ogden},
  {Ordenovic}, {Osinde}, {Pagani}, {Pagano}, {Palaversa}, {Palicio},
  {Pallas-Quintela}, {Panahi}, {Payne-Wardenaar}, {Pe{\~n}alosa Esteller},
  {Penttil{\"a}}, {Pichon}, {Piersimoni}, {Pineau}, {Plachy}, {Plum}, {Poggio},
  {Pr{\v{s}}a}, {Pulone}, {Racero}, {Ragaini}, {Rainer}, {Raiteri}, {Rambaux},
  {Ramos}, {Ramos-Lerate}, {Re Fiorentin}, {Regibo}, {Richards}, {Rios Diaz},
  {Ripepi}, {Riva}, {Rix}, {Rixon}, {Robichon}, {Robin}, {Robin}, {Roelens},
  {Rogues}, {Rohrbasser}, {Romero-G{\'o}mez}, {Rowell}, {Royer}, {Ruz Mieres},
  {Rybicki}, {Sadowski}, {S{\'a}ez N{\'u}{\~n}ez}, {Sagrist{\`a} Sell{\'e}s},
  {Sahlmann}, {Salguero}, {Samaras}, {Sanchez Gimenez}, {Sanna},
  {Santove{\~n}a}, {Sarasso}, {Schultheis}, {Sciacca}, {Segol}, {Segovia},
  {S{\'e}gransan}, {Semeux}, {Shahaf}, {Siddiqui}, {Siebert}, {Siltala},
  {Silvelo}, {Slezak}, {Slezak}, {Smart}, {Snaith}, {Solano}, {Solitro},
  {Souami}, {Souchay}, {Spagna}, {Spina}, {Spoto}, {Steele},
  {Steidelm{\"u}ller}, {Stephenson}, {S{\"u}veges}, {Surdej}, {Szabados},
  {Szegedi-Elek}, {Taris}, {Taylor}, {Teixeira}, {Tolomei}, {Tonello}, {Torra},
  {Torra}, {Torralba Elipe}, {Trabucchi}, {Tsounis}, {Turon}, {Ulla}, {Unger},
  {Vaillant}, {van Dillen}, {van Reeven}, {Vanel}, {Vecchiato}, {Viala},
  {Vicente}, {Voutsinas}, {Weiler}, {Wevers}, {Wyrzykowski}, {Yoldas}, {Yvard},
  {Zhao}, {Zorec}, {Zucker}, \& {Zwitter}}]{gaia2023A&A}
{Gaia Collaboration}, {Vallenari}, A., {Brown}, A.~G.~A., {et~al.} 2023, \aap,
  674, A1, \dodoi{10.1051/0004-6361/202243940}

\bibitem[{{Gallart} {et~al.}(2019){Gallart}, {Bernard}, {Brook}, {Ruiz-Lara},
  {Cassisi}, {Hill}, \& {Monelli}}]{Gallart2019NatAs}
{Gallart}, C., {Bernard}, E.~J., {Brook}, C.~B., {et~al.} 2019, Nature
  Astronomy, 3, 932, \dodoi{10.1038/s41550-019-0829-5}

\bibitem[{{Grand} {et~al.}(2017){Grand}, {G{\'o}mez}, {Marinacci}, {Pakmor},
  {Springel}, {Campbell}, {Frenk}, {Jenkins}, \& {White}}]{Grand2017}
{Grand}, R. J.~J., {G{\'o}mez}, F.~A., {Marinacci}, F., {et~al.} 2017, \mnras,
  467, 179, \dodoi{10.1093/mnras/stx071}

\bibitem[{{GRAVITY Collaboration} {et~al.}(2018){GRAVITY Collaboration},
  {Abuter}, {Amorim}, {Anugu}, {Baub{\"o}ck}, {Benisty}, {Berger}, {Blind},
  {Bonnet}, {Brandner}, {Buron}, {Collin}, {Chapron}, {Cl{\'e}net}, {Coud{\'e}
  Du Foresto}, {de Zeeuw}, {Deen}, {Delplancke-Str{\"o}bele}, {Dembet},
  {Dexter}, {Duvert}, {Eckart}, {Eisenhauer}, {Finger}, {F{\"o}rster
  Schreiber}, {F{\'e}dou}, {Garcia}, {Garcia Lopez}, {Gao}, {Gendron},
  {Genzel}, {Gillessen}, {Gordo}, {Habibi}, {Haubois}, {Haug}, {Hau{\ss}mann},
  {Henning}, {Hippler}, {Horrobin}, {Hubert}, {Hubin}, {Jimenez Rosales},
  {Jochum}, {Jocou}, {Kaufer}, {Kellner}, {Kendrew}, {Kervella}, {Kok},
  {Kulas}, {Lacour}, {Lapeyr{\`e}re}, {Lazareff}, {Le Bouquin}, {L{\'e}na},
  {Lippa}, {Lenzen}, {M{\'e}rand}, {M{\"u}ler}, {Neumann}, {Ott}, {Palanca},
  {Paumard}, {Pasquini}, {Perraut}, {Perrin}, {Pfuhl}, {Plewa}, {Rabien},
  {Ram{\'\i}rez}, {Ramos}, {Rau}, {Rodr{\'\i}guez-Coira}, {Rohloff}, {Rousset},
  {Sanchez-Bermudez}, {Scheithauer}, {Sch{\"o}ller}, {Schuler}, {Spyromilio},
  {Straub}, {Straubmeier}, {Sturm}, {Tacconi}, {Tristram}, {Vincent}, {von
  Fellenberg}, {Wank}, {Waisberg}, {Widmann}, {Wieprecht}, {Wiest},
  {Wiezorrek}, {Woillez}, {Yazici}, {Ziegler}, \&
  {Zins}}]{GRAVITYCollaboration2018A&A}
{GRAVITY Collaboration}, {Abuter}, R., {Amorim}, A., {et~al.} 2018, \aap, 615,
  L15, \dodoi{10.1051/0004-6361/201833718}

\bibitem[{Hasselquist {et~al.}(2021)Hasselquist, Hayes, Lian, Weinberg,
  Zasowski, Horta, Beaton, Feuillet, Garro, Gallart, Smith, Holtzman, Minniti,
  Lacerna, Shetrone, Jönsson, Cioni, Fillingham, Cunha, O’Connell,
  Fernández-Trincado, Muñoz, Schiavon, Almeida, Anguiano, Beers, Bizyaev,
  Brownstein, Cohen, Frinchaboy, García-Hernández, Geisler, Lane, Majewski,
  Nidever, Nitschelm, Povick, Price-Whelan, Roman-Lopes, Rosado, Sobeck,
  Stringfellow, Valenzuela, Villanova, \& Vincenzo}]{Hasselquist2021}
Hasselquist, S., Hayes, C.~R., Lian, J., {et~al.} 2021, \apj, 923, 172,
  \dodoi{10.3847/1538-4357/ac25f9}

\bibitem[{{Hawkins} {et~al.}(2015){Hawkins}, {Jofr{\'e}}, {Masseron}, \&
  {Gilmore}}]{Hawkins2015using}
{Hawkins}, K., {Jofr{\'e}}, P., {Masseron}, T., \& {Gilmore}, G. 2015, \mnras,
  453, 758, \dodoi{10.1093/mnras/stv1586}

\bibitem[{{Haywood} {et~al.}(2018){Haywood}, {Di Matteo}, {Lehnert}, {Snaith},
  {Khoperskov}, \& {G{\'o}mez}}]{Haywood2018in}
{Haywood}, M., {Di Matteo}, P., {Lehnert}, M.~D., {et~al.} 2018, \apj, 863,
  113, \dodoi{10.3847/1538-4357/aad235}

\bibitem[{{Helmi} {et~al.}(2018){Helmi}, {Babusiaux}, {Koppelman}, {Massari},
  {Veljanoski}, \& {Brown}}]{Helmi2018the}
{Helmi}, A., {Babusiaux}, C., {Koppelman}, H.~H., {et~al.} 2018, \nat, 563, 85,
  \dodoi{10.1038/s41586-018-0625-x}

\bibitem[{{Helmi} \& {de Zeeuw}(2000)}]{HelmiZeeuw2000MNRAS}
{Helmi}, A., \& {de Zeeuw}, P.~T. 2000, \mnras, 319, 657,
  \dodoi{10.1046/j.1365-8711.2000.03895.x}

\bibitem[{{Helmi} {et~al.}(1999){Helmi}, {White}, {de Zeeuw}, \&
  {Zhao}}]{Helmi1999Debris}
{Helmi}, A., {White}, S. D.~M., {de Zeeuw}, P.~T., \& {Zhao}, H. 1999, \nat,
  402, 53, \dodoi{10.1038/46980}

\bibitem[{{Horta} {et~al.}(2021){Horta}, {Schiavon}, {Mackereth}, {Pfeffer},
  {Mason}, {Kisku}, {Fragkoudi}, {Allende Prieto}, {Cunha}, {Hasselquist},
  {Holtzman}, {Majewski}, {Nataf}, {O'Connell}, {Schultheis}, \&
  {Smith}}]{Horta2021evidence}
{Horta}, D., {Schiavon}, R.~P., {Mackereth}, J.~T., {et~al.} 2021, \mnras, 500,
  1385, \dodoi{10.1093/mnras/staa2987}

\bibitem[{{Horta} {et~al.}(2023){Horta}, {Schiavon}, {Mackereth}, {Weinberg},
  {Hasselquist}, {Feuillet}, {O'Connell}, {Anguiano}, {Allende-Prieto},
  {Beaton}, {Bizyaev}, {Cunha}, {Geisler}, {Garc{\'\i}a-Hern{\'a}ndez},
  {Holtzman}, {J{\"o}nsson}, {Lane}, {Majewski}, {M{\'e}sz{\'a}ros}, {Minniti},
  {Nitschelm}, {Shetrone}, {Smith}, \& {Zasowski}}]{Horta2023the}
---. 2023, \mnras, 520, 5671, \dodoi{10.1093/mnras/stac3179}

\bibitem[{{Ibata} {et~al.}(2020){Ibata}, {Bellazzini}, {Thomas}, {Malhan},
  {Martin}, {Famaey}, \& {Siebert}}]{Ibata2020apl}
{Ibata}, R., {Bellazzini}, M., {Thomas}, G., {et~al.} 2020, \apjl, 891, L19,
  \dodoi{10.3847/2041-8213/ab77c7}

\bibitem[{{Ibata} {et~al.}(2001){Ibata}, {Irwin}, {Lewis}, \&
  {Stolte}}]{Ibata2001galactic}
{Ibata}, R., {Irwin}, M., {Lewis}, G.~F., \& {Stolte}, A. 2001, \apjl, 547,
  L133, \dodoi{10.1086/318894}

\bibitem[{{Ibata} {et~al.}(1994){Ibata}, {Gilmore}, \& {Irwin}}]{Ibata1994ads}
{Ibata}, R.~A., {Gilmore}, G., \& {Irwin}, M.~J. 1994, \nat, 370, 194,
  \dodoi{10.1038/370194a0}

\bibitem[{{Jean-Baptiste} {et~al.}(2017){Jean-Baptiste}, {Di Matteo},
  {Haywood}, {G{\'o}mez}, {Montuori}, {Combes}, \&
  {Semelin}}]{Jean-Baptiste2017onJ}
{Jean-Baptiste}, I., {Di Matteo}, P., {Haywood}, M., {et~al.} 2017, \aap, 604,
  A106, \dodoi{10.1051/0004-6361/201629691}

\bibitem[{{Khoperskov} {et~al.}(2023){Khoperskov}, {Minchev}, {Libeskind},
  {Haywood}, {Di Matteo}, {Belokurov}, {Steinmetz}, {Gomez}, {Grand},
  {Hoffman}, {Knebe}, {Sorce}, {Spaare}, {Tempel}, \&
  {Vogelsberger}}]{Khoperskov2023A&A}
{Khoperskov}, S., {Minchev}, I., {Libeskind}, N., {et~al.} 2023, \aap, 677,
  A90, \dodoi{10.1051/0004-6361/202244233}

\bibitem[{{Kilic} {et~al.}(2017){Kilic}, {Munn}, {Harris}, {von Hippel},
  {Liebert}, {Williams}, {Jeffery}, \& {DeGennaro}}]{Kilic2017the}
{Kilic}, M., {Munn}, J.~A., {Harris}, H.~C., {et~al.} 2017, \apj, 837, 162,
  \dodoi{10.3847/1538-4357/aa62a5}

\bibitem[{Koppelman {et~al.}(2018)Koppelman, Helmi, \&
  Veljanoski}]{Koppelman_2018}
Koppelman, H., Helmi, A., \& Veljanoski, J. 2018, \apjl, 860, L11,
  \dodoi{10.3847/2041-8213/aac882}

\bibitem[{{Koppelman} {et~al.}(2019{\natexlab{a}}){Koppelman}, {Helmi},
  {Massari}, {Price-Whelan}, \& {Starkenburg}}]{Koppelman2019Multiple}
{Koppelman}, H.~H., {Helmi}, A., {Massari}, D., {Price-Whelan}, A.~M., \&
  {Starkenburg}, T.~K. 2019{\natexlab{a}}, \aap, 631, L9,
  \dodoi{10.1051/0004-6361/201936738}

\bibitem[{{Koppelman} {et~al.}(2019{\natexlab{b}}){Koppelman}, {Helmi},
  {Massari}, {Roelenga}, \& {Bastian}}]{Koppelman2019characterization}
{Koppelman}, H.~H., {Helmi}, A., {Massari}, D., {Roelenga}, S., \& {Bastian},
  U. 2019{\natexlab{b}}, \aap, 625, A5, \dodoi{10.1051/0004-6361/201834769}

\bibitem[{Leung \& Bovy(2018)}]{Leung20181}
Leung, H.~W., \& Bovy, J. 2018, \mnras, 483, 3255,
  \dodoi{10.1093/mnras/sty3217}

\bibitem[{{Leung} \& {Bovy}(2019)}]{Leung2019}
{Leung}, H.~W., \& {Bovy}, J. 2019, \mnras, 489, 2079,
  \dodoi{10.1093/mnras/stz2245}

\bibitem[{{Mackereth} \& {Bovy}(2018)}]{Mackereth2018PASP}
{Mackereth}, J.~T., \& {Bovy}, J. 2018, \pasp, 130, 114501,
  \dodoi{10.1088/1538-3873/aadcdd}

\bibitem[{{Mackereth} {et~al.}(2019{\natexlab{a}}){Mackereth}, {Bovy}, {Leung},
  {Schiavon}, {Trick}, {Chaplin}, {Cunha}, {Feuillet}, {Majewski}, {Martig},
  {Miglio}, {Nidever}, {Pinsonneault}, {Aguirre}, {Sobeck}, {Tayar}, \&
  {Zasowski}}]{Mackereth2019}
{Mackereth}, J.~T., {Bovy}, J., {Leung}, H.~W., {et~al.} 2019{\natexlab{a}},
  \mnras, 489, 176, \dodoi{10.1093/mnras/stz1521}

\bibitem[{{Mackereth} {et~al.}(2019{\natexlab{b}}){Mackereth}, {Schiavon},
  {Pfeffer}, {Hayes}, {Bovy}, {Anguiano}, {Allende Prieto}, {Hasselquist},
  {Holtzman}, {Johnson}, {Majewski}, {O'Connell}, {Shetrone}, {Tissera}, \&
  {Fern{\'a}ndez-Trincado}}]{Mackereth2019the}
{Mackereth}, J.~T., {Schiavon}, R.~P., {Pfeffer}, J., {et~al.}
  2019{\natexlab{b}}, \mnras, 482, 3426, \dodoi{10.1093/mnras/sty2955}

\bibitem[{{Majewski} {et~al.}(2003){Majewski}, {Skrutskie}, {Weinberg}, \&
  {Ostheimer}}]{Majewski2003atm}
{Majewski}, S.~R., {Skrutskie}, M.~F., {Weinberg}, M.~D., \& {Ostheimer}, J.~C.
  2003, \apj, 599, 1082, \dodoi{10.1086/379504}

\bibitem[{{Malhan} {et~al.}(2022){Malhan}, {Ibata}, {Sharma}, {Famaey},
  {Bellazzini}, {Carlberg}, {D'Souza}, {Yuan}, {Martin}, \&
  {Thomas}}]{Malhan2022the}
{Malhan}, K., {Ibata}, R.~A., {Sharma}, S., {et~al.} 2022, \apj, 926, 107,
  \dodoi{10.3847/1538-4357/ac4d2a}

\bibitem[{{Matsuno} {et~al.}(2019){Matsuno}, {Aoki}, \&
  {Suda}}]{Matsuno2019origin}
{Matsuno}, T., {Aoki}, W., \& {Suda}, T. 2019, \apjl, 874, L35,
  \dodoi{10.3847/2041-8213/ab0ec0}

\bibitem[{{Mori} {et~al.}(2024){Mori}, {Di Matteo}, {Salvadori}, {Khoperskov},
  {Pagnini}, \& {Haywood}}]{Mori2024eas}
{Mori}, A., {Di Matteo}, P., {Salvadori}, S., {et~al.} 2024, \aap, 690, A136,
  \dodoi{10.1051/0004-6361/202449291}

\bibitem[{{Myeong} {et~al.}(2018){Myeong}, {Evans}, {Belokurov}, {Sanders}, \&
  {Koposov}}]{Myeong2018the}
{Myeong}, G.~C., {Evans}, N.~W., {Belokurov}, V., {Sanders}, J.~L., \&
  {Koposov}, S.~E. 2018, \apjl, 863, L28, \dodoi{10.3847/2041-8213/aad7f7}

\bibitem[{{Myeong} {et~al.}(2019){Myeong}, {Vasiliev}, {Iorio}, {Evans}, \&
  {Belokurov}}]{Myeong2019evidence}
{Myeong}, G.~C., {Vasiliev}, E., {Iorio}, G., {Evans}, N.~W., \& {Belokurov},
  V. 2019, \mnras, 488, 1235, \dodoi{10.1093/mnras/stz1770}

\bibitem[{{Naidu} {et~al.}(2020){Naidu}, {Conroy}, {Bonaca}, {Johnson}, {Ting},
  {Caldwell}, {Zaritsky}, \& {Cargile}}]{Naidu2020evidence}
{Naidu}, R.~P., {Conroy}, C., {Bonaca}, A., {et~al.} 2020, \apj, 901, 48,
  \dodoi{10.3847/1538-4357/abaef4}

\bibitem[{{Necib} {et~al.}(2020){Necib}, {Ostdiek}, {Lisanti}, {Cohen},
  {Freytsis}, {Garrison-Kimmel}, {Hopkins}, {Wetzel}, \&
  {Sanderson}}]{Necib2020evidence}
{Necib}, L., {Ostdiek}, B., {Lisanti}, M., {et~al.} 2020, Nature Astronomy, 4,
  1078, \dodoi{10.1038/s41550-020-1131-2}

\bibitem[{{Newberg} {et~al.}(2009){Newberg}, {Yanny}, \&
  {Willett}}]{Newberg2009discovery}
{Newberg}, H.~J., {Yanny}, B., \& {Willett}, B.~A. 2009, \apjl, 700, L61,
  \dodoi{10.1088/0004-637X/700/2/L61}

\bibitem[{{Pagnini} {et~al.}(2023){Pagnini}, {Di Matteo}, {Khoperskov},
  {Mastrobuono-Battisti}, {Haywood}, {Renaud}, \& {Combes}}]{Pagnini2023A&A}
{Pagnini}, G., {Di Matteo}, P., {Khoperskov}, S., {et~al.} 2023, \aap, 673,
  A86, \dodoi{10.1051/0004-6361/202245128}

\bibitem[{{Quinn} \& {Goodman}(1986)}]{Quinn1986ApJ}
{Quinn}, P.~J., \& {Goodman}, J. 1986, \apj, 309, 472, \dodoi{10.1086/164619}

\bibitem[{{Re Fiorentin} {et~al.}(2021){Re Fiorentin}, {Spagna}, {Lattanzi}, \&
  {Cignoni}}]{ReFiorentin2021Icarus}
{Re Fiorentin}, P., {Spagna}, A., {Lattanzi}, M.~G., \& {Cignoni}, M. 2021,
  \apjl, 907, L16, \dodoi{10.3847/2041-8213/abd53d}

\bibitem[{{Schiavon} {et~al.}(2024){Schiavon}, {Phillips}, {Myers}, {Horta},
  {Minniti}, {Allende Prieto}, {Anguiano}, {Beaton}, {Beers}, {Brownstein},
  {Cohen}, {Fern{\'a}ndez-Trincado}, {Frinchaboy}, {J{\"o}nsson}, {Kisku},
  {Lane}, {Majewski}, {Mason}, {M{\'e}sz{\'a}ros}, \&
  {Stringfellow}}]{Schiavon2024}
{Schiavon}, R.~P., {Phillips}, S.~G., {Myers}, N., {et~al.} 2024, \mnras, 528,
  1393, \dodoi{10.1093/mnras/stad3020}

\bibitem[{{Sch{\"o}nrich} {et~al.}(2010){Sch{\"o}nrich}, {Binney}, \&
  {Dehnen}}]{sch2010MNRAS}
{Sch{\"o}nrich}, R., {Binney}, J., \& {Dehnen}, W. 2010, \mnras, 403, 1829,
  \dodoi{10.1111/j.1365-2966.2010.16253.x}

\bibitem[{{Steinmetz} {et~al.}(2020){Steinmetz}, {Guiglion}, {McMillan},
  {Matijevi{\v{c}}}, {Enke}, {Kordopatis}, {Zwitter}, {Valentini}, {Chiappini},
  {Casagrande}, {Wojno}, {Anguiano}, {Bienaym{\'e}}, {Bijaoui}, {Binney},
  {Burton}, {Cass}, {de Laverny}, {Fiegert}, {Freeman}, {Fulbright}, {Gibson},
  {Gilmore}, {Grebel}, {Helmi}, {Kunder}, {Munari}, {Navarro}, {Parker},
  {Ruchti}, {Recio-Blanco}, {Reid}, {Seabroke}, {Siviero}, {Siebert}, {Stupar},
  {Watson}, {Williams}, {Wyse}, {Anders}, {Antoja}, {Birko}, {Bland-Hawthorn},
  {Bossini}, {Garc{\'\i}a}, {Carrillo}, {Chaplin}, {Elsworth}, {Famaey},
  {Gerhard}, {Jofre}, {Just}, {Mathur}, {Miglio}, {Minchev}, {Monari},
  {Mosser}, {Ritter}, {Rodrigues}, {Scholz}, {Sharma}, {Sysoliatina}, \& {RAVE
  Collaboration}}]{Steinmetz2020the}
{Steinmetz}, M., {Guiglion}, G., {McMillan}, P.~J., {et~al.} 2020, \aj, 160,
  83, \dodoi{10.3847/1538-3881/ab9ab8}

\bibitem[{{Timmes} {et~al.}(1995){Timmes}, {Woosley}, \&
  {Weaver}}]{Timmes1995ApJS}
{Timmes}, F.~X., {Woosley}, S.~E., \& {Weaver}, T.~A. 1995, \apjs, 98, 617,
  \dodoi{10.1086/192172}

\bibitem[{Tononi {et~al.}(2019)Tononi, Torres, García-Berro, Camisassa,
  Althaus, \& Rebassa-Mansergas}]{Tononi2019effects}
Tononi, J., Torres, S., García-Berro, E., {et~al.} 2019, \aap, 628, A52,
  \dodoi{10.1051/0004-6361/201834267}

\bibitem[{{Vasiliev} \& {Belokurov}(2020)}]{Vasiliev2020the}
{Vasiliev}, E., \& {Belokurov}, V. 2020, \mnras, 497, 4162,
  \dodoi{10.1093/mnras/staa2114}

\bibitem[{{Walker} {et~al.}(1996){Walker}, {Mihos}, \&
  {Hernquist}}]{Walker1996ApJ}
{Walker}, I.~R., {Mihos}, J.~C., \& {Hernquist}, L. 1996, \apj, 460, 121,
  \dodoi{10.1086/176956}

\bibitem[{{Wang} {et~al.}(2023){Wang}, {Necib}, {Ji}, {Ou}, {Lisanti}, {de los
  Reyes}, {Strom}, \& {Truong}}]{Wang2023ApJ}
{Wang}, S., {Necib}, L., {Ji}, A.~P., {et~al.} 2023, \apj, 955, 129,
  \dodoi{10.3847/1538-4357/acec4d}

\bibitem[{{White} \& {Frenk}(1991)}]{White1991ApJ}
{White}, S. D.~M., \& {Frenk}, C.~S. 1991, \apj, 379, 52,
  \dodoi{10.1086/170483}

\bibitem[{{Wu} {et~al.}(2022){Wu}, {Zhao}, {Xue}, {Bird}, \&
  {Yang}}]{Wu2022contribution}
{Wu}, W., {Zhao}, G., {Xue}, X.-X., {Bird}, S.~A., \& {Yang}, C. 2022, \apj,
  924, 23, \dodoi{10.3847/1538-4357/ac31ac}

\bibitem[{{Yanny} {et~al.}(2009){Yanny}, {Rockosi}, {Newberg}, {Knapp},
  {Adelman-McCarthy}, {Alcorn}, {Allam}, {Allende Prieto}, {An}, {Anderson},
  {Anderson}, {Bailer-Jones}, {Bastian}, {Beers}, {Bell}, {Belokurov},
  {Bizyaev}, {Blythe}, {Bochanski}, {Boroski}, {Brinchmann}, {Brinkmann},
  {Brewington}, {Carey}, {Cudworth}, {Evans}, {Evans}, {Gates}, {G{\"a}nsicke},
  {Gillespie}, {Gilmore}, {Nebot Gomez-Moran}, {Grebel}, {Greenwell}, {Gunn},
  {Jordan}, {Jordan}, {Harding}, {Harris}, {Hendry}, {Holder}, {Ivans},
  {Ivezi{\v{c}}}, {Jester}, {Johnson}, {Kent}, {Kleinman}, {Kniazev},
  {Krzesinski}, {Kron}, {Kuropatkin}, {Lebedeva}, {Lee}, {French Leger},
  {L{\'e}pine}, {Levine}, {Lin}, {Long}, {Loomis}, {Lupton}, {Malanushenko},
  {Malanushenko}, {Margon}, {Martinez-Delgado}, {McGehee}, {Monet}, {Morrison},
  {Munn}, {Neilsen}, {Nitta}, {Norris}, {Oravetz}, {Owen}, {Padmanabhan},
  {Pan}, {Peterson}, {Pier}, {Platson}, {Re Fiorentin}, {Richards}, {Rix},
  {Schlegel}, {Schneider}, {Schreiber}, {Schwope}, {Sibley}, {Simmons},
  {Snedden}, {Allyn Smith}, {Stark}, {Stauffer}, {Steinmetz}, {Stoughton},
  {SubbaRao}, {Szalay}, {Szkody}, {Thakar}, {Sivarani}, {Tucker}, {Uomoto},
  {Vanden Berk}, {Vidrih}, {Wadadekar}, {Watters}, {Wilhelm}, {Wyse}, {Yarger},
  \& {Zucker}}]{Yanny2009segue}
{Yanny}, B., {Rockosi}, C., {Newberg}, H.~J., {et~al.} 2009, \aj, 137, 4377,
  \dodoi{10.1088/0004-6256/137/5/4377}

\bibitem[{{Ye} {et~al.}(2024{\natexlab{a}}){Ye}, {Du}, {Deng}, {Liao}, {Huang},
  {Shi}, \& {Ma}}]{Ye2024compositions}
{Ye}, D., {Du}, C., {Deng}, M., {et~al.} 2024{\natexlab{a}}, \mnras, 532, 2584,
  \dodoi{10.1093/mnras/stae1655}

\bibitem[{{Ye} {et~al.}(2024{\natexlab{b}}){Ye}, {Du}, {Shi}, \&
  {Ma}}]{Ye2024dynamical}
{Ye}, D., {Du}, C., {Shi}, J., \& {Ma}, J. 2024{\natexlab{b}}, \mnras, 527,
  9892, \dodoi{10.1093/mnras/stad3860}

\bibitem[{{Yuan} {et~al.}(2018){Yuan}, {Chang}, {Banerjee}, {Han}, {Kang}, \&
  {Smith}}]{Yuan2018StarGO}
{Yuan}, Z., {Chang}, J., {Banerjee}, P., {et~al.} 2018, \apj, 863, 26,
  \dodoi{10.3847/1538-4357/aacd0d}

\bibitem[{{Yuan} {et~al.}(2020{\natexlab{a}}){Yuan}, {Chang}, {Beers}, \&
  {Huang}}]{Yuan2020als}
{Yuan}, Z., {Chang}, J., {Beers}, T.~C., \& {Huang}, Y. 2020{\natexlab{a}},
  \apjl, 898, L37, \dodoi{10.3847/2041-8213/aba49f}

\bibitem[{{Yuan} {et~al.}(2019){Yuan}, {Smith}, {Xue}, {Li}, {Liu}, {Wang},
  {Li}, \& {Chang}}]{Yuan2019revealing}
{Yuan}, Z., {Smith}, M.~C., {Xue}, X.-X., {et~al.} 2019, \apj, 881, 164,
  \dodoi{10.3847/1538-4357/ab2e09}

\bibitem[{{Yuan} {et~al.}(2020{\natexlab{b}}){Yuan}, {Myeong}, {Beers},
  {Evans}, {Lee}, {Banerjee}, {Gudin}, {Hattori}, {Li}, {Matsuno}, {Placco},
  {Smith}, {Whitten}, \& {Zhao}}]{Yuan2020dynamic}
{Yuan}, Z., {Myeong}, G.~C., {Beers}, T.~C., {et~al.} 2020{\natexlab{b}}, \apj,
  891, 39, \dodoi{10.3847/1538-4357/ab6ef7}

\bibitem[{Yuan {et~al.}(2022)Yuan, Malhan, Sestito, Ibata, Martin, Chang, Li,
  Caffau, Bonifacio, Bellazzini, Huang, Voggel, Longeard, Arentsen,
  Doliva-Dolinsky, Navarro, Famaey, Starkenburg, \& Aguado}]{Yuan2022the}
Yuan, Z., Malhan, K., Sestito, F., {et~al.} 2022, \apj, 930, 103,
  \dodoi{10.3847/1538-4357/ac616f}

\bibitem[{{Zhao} {et~al.}(2012){Zhao}, {Zhao}, {Chu}, {Jing}, \&
  {Deng}}]{Zhao2012RAA}
{Zhao}, G., {Zhao}, Y.-H., {Chu}, Y.-Q., {Jing}, Y.-P., \& {Deng}, L.-C. 2012,
  RAA, 12, 723, \dodoi{10.1088/1674-4527/12/7/002}

\bibitem[{{Zucker} {et~al.}(2021){Zucker}, {Simpson}, {Martell}, {Lewis},
  {Casey}, {Ting}, {Horner}, {Nordlander}, {Wyse}, {Zwitter}, {Bland-Hawthorn},
  {Buder}, {Asplund}, {De Silva}, {D'Orazi}, {Freeman}, {Hayden}, {Kos}, {Lin},
  {Lind}, {Schlesinger}, {Sharma}, \& {Stello}}]{Zucker2021ApJ}
{Zucker}, D.~B., {Simpson}, J.~D., {Martell}, S.~L., {et~al.} 2021, \apjl, 912,
  L30, \dodoi{10.3847/2041-8213/abf7cd}

\end{thebibliography}
\bibliographystyle{aasjournal}



\end{document}